\documentclass{acmart}

\AtBeginDocument{%
  \providecommand\BibTeX{{%
    \normalfont B\kern-0.5em{\scshape i\kern-0.25em b}\kern-0.8em\TeX}}}

\copyrightyear{2022}
\acmYear{2022}
\setcopyright{acmlicensed}\acmConference[FAccT '22]{2022 ACM Conference on Fairness, Accountability, and Transparency}{June 21--24, 2022}{Seoul, Republic of Korea}
\acmBooktitle{2022 ACM Conference on Fairness, Accountability, and Transparency (FAccT '22), June 21--24, 2022, Seoul, Republic of Korea}
\acmPrice{15.00}
\acmDOI{10.1145/3531146.3533169}
\acmISBN{978-1-4503-9352-2/22/06}

\begin{document}

\title[Tackling Algorithmic Disability Discrimination in the Hiring Process]
{Tackling Algorithmic Disability Discrimination in the Hiring Process: An Ethical, Legal and Technical Analysis}%

\author{Maarten Buyl}
\authornote{All authors contributed equally to this research.}
\affiliation{%
  \institution{Ghent University}
  \country{Belgium}
}
\email{maarten.buyl@ugent.be}
\orcid{0000-0002-5434-2386}

\author{Christina Cociancig}
\authornotemark[1]
\affiliation{%
  \institution{University of Bremen}
  \country{Germany}
}
\email{chrcoc@uni-bremen.de}
\orcid{0000-0002-5645-6432}

\author{Cristina Frattone}
\authornotemark[1]
\affiliation{%
  \institution{Roma Tre University}
  \country{Italy}
}
\email{cristina.frattone@uniroma3.it}
\orcid{0000-0003-1734-243X}

\author{Nele Roekens}
\authornotemark[1]
\affiliation{%
  \institution{Unia}
  \country{Belgium}
}
\email{nele.roekens@unia.be}
\orcid{0000-0002-1975-1758}


\begin{abstract}
Tackling algorithmic discrimination against persons with disabilities (PWDs) demands a distinctive approach that is fundamentally different to that applied to other protected characteristics, due to particular ethical, legal, and technical challenges. We address these challenges specifically in the context of artificial intelligence (AI) systems used in hiring processes (or automated hiring systems, AHSs), in which automated assessment procedures are subject to unique ethical and legal considerations and have an undeniable adverse impact on PWDs. In this paper, we discuss concerns and opportunities raised by AI-driven hiring in relation to disability discrimination. Ultimately, we aim to encourage further research into this topic. Hence, we establish some starting points and design a roadmap for ethicists, lawmakers, advocates as well as AI practitioners alike.
\end{abstract}



\maketitle
\section{Introduction}
Recent advancements in computer science have made \textit{artificial intelligence} (AI) more attractive than ever for both private and public actors. As a result, AI is highly pervasive nowadays; indeed it is difficult to find any aspect of everyday life that is not affected by it. Inspired by the proposed EU Regulation on AI (AIA) \cite{AIA2021}, we consider AI as software used to generate predictions, recommendations or decisions, developed with machine learning or other statistical methods. To draw statistical inferences, AI systems are “trained” with the aim of generating predictions that are as accurate as possible. Yet, when predictions affect the lives of individuals, competing interests about the optimization of design and predictions take over \cite{barocas2016big}.

In particular, we study the use of AI as part of automated hiring systems (AHSs), with the help of which employers want to attract and hire the best employees with the lowest costs. Prospective employees, however, ought to be protected from unjust discrimination. AI systems should therefore not only strive to predict which employees are a good fit for a job vacancy, but also strive to avoid \textit{algorithmic discrimination} or \textit{algorithmic bias}. This form of discrimination or bias describes the unjust discriminatory manner in which (partially or fully) automated decisions impact a certain protected group or minority. Discrimination can be caused, e.g., by a disproportionate distribution of prediction errors, a faulty design of the AI architecture, or a historical bias rooted in society and perpetuated by statistical inferences \cite{mehrabi2021survey, hacker2018teaching}. Consequently, in many ways algorithmic bias is the perpetuation of \textit{human} bias. 

Our specific concern is for the algorithmic discrimination experienced by persons with disabilities (PWDs)\footnote{Considering the importance of language in writing about PWDs, we respect the guidelines laid out in \cite{adannguideline}, which recommend referring to the person first and the disability second.} during hiring processes. In a recent study \cite{sanchez2020does} of three AHSs, algorithmic bias was actively being reduced for some protected traits, yet disability was not mentioned in the examples nor in the validation tests. Even though PWDs find no mention in studies like this, it has been recognized that they face unique forms of discrimination in their everyday life, including when they apply for a job \cite{olkin2002could}. With the aim of inspiring further research into this area, we provide an analysis of the ethical, legal and technical challenges and opportunities for promoting social justice for job candidates with a disability.

First in Section \ref{sec:ethical}, we argue that non-discrimination and inclusiveness should be regarded as moral obligations to a distinguishable vulnerable group. We identify and examine the importance of ethics of information, including the decision for or against disclosure of a disability. Concerning automated decision making, we refer to AI Ethics publications. A quantitative study of 84 comprehensive guidelines supports our findings of discriminatory practices towards PWDs. We finally bridge the ethical with the legal perspective, establish commonalities and draw conclusions. 

Second in Section \ref{sec:legal}, we focus on the relevant EU legal framework. Notably, EU equality law \cite{EED2000}, which prescribes \textit{reasonable accommodation} for PWDs in hiring procedures, including automated processes. We maintain that failure to provide reasonable accommodation to a PWD by an AHS amounts to \textit{prima facie} discrimination, thus shifting the burden of proof to the defendant. However, while adequate accommodation may require disclosure of disability, the processing of disability data for this purpose needs further legal clarification. Finally, we discuss the challenges and opportunities that stem from the rights and obligations provided by the General Data Protection Regulation (GDPR) \cite{GDPR2016} and the AIA \cite{AIA2021}.

Third in Section \ref{sec:technical}, we present technical challenges and possible avenues of research. As each disability is particular and the existing legal framework is applied on a case-by-case basis, data-driven approaches show little promise to solve discrimination by AHSs. AI would be expected to make problematic assumptions, including about the limitations of each PWD, the necessity of an accommodation, and its degree of reasonableness. Finally, we highlight the opportunities that AI systems present for the assessment of PWDs, such as recommending the extent of accommodations and proving discrimination. 

\section{Ethical Analysis}\label{sec:ethical}
As a first step we consider the ethical perspective of tackling algorithmic discrimination against PWDs. On the question of morality, it is evident that as a society we need to value non-discrimination and inclusion of minority groups and ensure equal opportunities. However, we aim to raise this value to the level of a moral obligation towards PWDs. In this section, we first lay out arguments for such an amplification in Section \ref{sec:moral}. Moreover, we introduce our findings regarding the ethics of AI specifically in Section \ref{sec:aiethics} and finally define the role of information for non-discrimination of PWDs in Section \ref{sec:eth_info}. We conclude by attempting to develop an interdisciplinary bridge from ethics to law in Section \ref{sec:bridge}.

\subsection{Non-discrimination and inclusion as moral obligations}\label{sec:moral}
While anti discrimination law and societies' values agree on non-discrimination and inclusiveness of minority groups, it has not yet been engraved into our collective moral system as a proactive obligation. As \cite{moralimperative} demonstrates and arguments presented in Section \ref{sec:technical} support, individual case rulings and verdicts are insufficient to change the living experience of PWDs in the long term, even though they are compliant with law. It is time to initiate a shift in the morality of society towards an inclusive and non-discriminatory workplace.

We consider PWDs as a distinguishable vulnerable group among other minority groups. The following characteristics shortly outlined in \cite{olkin2002could} are specific to PWDs and provide a compelling argument for additional measures to protect their right to non-discrimination and inclusion:
\begin{enumerate}
\item PWDs experience separation and segregation distinct from other minority groups. With the intention of providing accessibility, PWDs are provided with separate entries, buses, bathrooms, classrooms, etc. As a result, PWDs are absent from “normal” classrooms, boardrooms, parliament etc. where diverse representation is needed.
\item PWDs mostly do not share their otherness with people in their surroundings, e.g., their families, resulting in an ill-equipped social environment to provide necessary accommodations.
\item Disabilities are often connected to pain, fatigue and muscle weakness. To manage bodily discomfort, costly resources like assistive technology are needed. 
\item PWDs have been perpetually discriminated against by means of language. For PWDs to move out of the sphere of “other”, they have made use of this instrument of oppression to reflect an ideological shift of perception from "the handicapped" to viewing the disability as a social construct instead of a defining fault.
\end{enumerate}
As a distinguishable vulnerable group, PWDs not only have the legal right to measures of accessibility, but it should be the society's moral imperative to provide those measures. Measures could include technical solutions as well as reasonable accommodation in the employment context. 

\subsection{Ethics of AI in relation to disabilities}\label{sec:aiethics}
As we focus on AI and the application thereof, we identified issues of discriminatory practices towards PWDs caused by automated decision making. This form of \textit{digital discrimination} reinforces the existing social inequality PWDs experience on a day-to-day basis \cite{aran2020bias}. AI describes a set of algorithms that rely heavily on finding patterns in high volumes of data. Patterns, for which those with disabilities once again fall outside of the area of applicability, as we examine from a technical perspective in Section \ref{sec:modeling}. This is just one effect of the datafication of humans that AI endorses. Unfortunately, it is easy to find examples of discrimination and questionable practices of number-crunching algorithms.

Even though technology can also assist and mitigate some disabilities, PWDs are affected negatively by AI in multiple areas and significantly more so than people without disabilities \cite{lillywhite2020coverage}. Affected areas are many, as AI is used in more and more versatile ways: humans are using AI to judge a PWD's qualification for an employment, loan, social benefits, etc. As a result, we see PWDs falling through the system more than ever. PWDs themselves use AI in many cases in therapeutic and non-therapeutic contexts. However, even in these contexts intended to make life easier, PWDs also experience discrimination, for example, in forms of voice recognition technology they cannot communicate with, because it simply does not recognize their voice as \textit{human} \cite{bariffi2021artificial}.

Due to their vulnerability as a social group, which is emphasized by the characteristics in Section \ref{sec:moral}, PWDs “face a greater risk of violation of their fundamental rights and freedoms, which justifies adopting specific approaches based on the principle of equality and non-discrimination” \cite{lillywhite2020coverage}. One would expect at least the AI Ethics community to ensure frameworks are protecting the fundamental rights of PWDs. However, as \cite{lillywhite2020coverage} recognized, PWDs are rarely mentioned in literature as anything other than therapeutic/non-therapeutic users of AI, with hardly any mentions of fairly processing PWDs with AHSs, and completely disregarding accessibility as well as the ability of PWDs to inform AI research or policies themselves. Repeatedly, PWDs are excluded and pushed into the shadows of society.

To support the findings of \cite{lillywhite2020coverage} for academic AI Ethics literature, we conducted a quantitative literary study of AI Ethics guidelines. We sampled AI Ethics publications listed in \cite{jobin2019global} and were able to successfully source 82 out of 84 listed guidelines. In these documents and websites, we conducted a keyword search for “disab*” to match “disabled”, “disability”, "people with disabilities" and similar phrases. The result provided support and motivation alike: only 22 out of 82 guidelines (26.8\%) mentioned “disabled people” or “people with disabilities” as a vulnerable group that should receive special consideration in AI Ethics guidelines. Furthermore, PWDs were not acknowledged as research members and policy makers of AI, despite their unique position and ability to shape future research into (assistive) technology and policy to their demands.

\subsection{Ethics of information}\label{sec:eth_info}
Most of the discussion surrounding discrimination of PWDs is related to information and not solely limited to the professional context and while applying AI. This includes the delicate decision for or against disclosure of a disability with the employer and the resulting tension of this decision, as well as the lack of information on both the employer’s and PWD’s sides concerning the rights surrounding reasonable accommodation. 

The decision for or against disclosure of a disability is motivated by the expectation of positive as well as negative outcomes of this decision. In a survey conducted in Canada \cite{gignac2021does} only 51\% of participants disclosed a disability at some point during their employment, basing their decision on an established secure and safe feeling within their work environment. Of those who decided against disclosure, most reported that they were able to manage their condition without it affecting their performance. However, 37\% did not feel secure enough to disclose their disability, and the same number reported that they were concerned about losing their employment. Numbers from Europe comparing unemployment rates of people with disabilities aged 20-64 (17.1\%) to people without disabilities (10.2\%) support this concern \cite{eupe651932}. Evidently, disclosure requires an appreciative and secure work environment that is built on trust of both employer and employee. 

To establish trust between employers and employees, legally and morally informed parties are required. For applicants and employees, this manifests itself as being able to disclose their disability as they see fit and because they feel secure enough to do so, as well as being enabled to claim accommodation where necessary. For the employer, this implies the moral obligation to establish a secure environment for applicants as well as employees, possibly even proactively offering accommodation, but certainly refraining from obstructing these processes. For both parties, information in the form of disclosure and legal knowledge should be available. Even though in some cases it has been argued that the interests of the employer overrule the right to privacy of an employee \cite{persson2003privacy}, we strongly argue against mandatory disclosure as it does not support the value of trust between parties nor a person's right to self-determination. 

\subsection{Bridging the ethical with the legal perspective}\label{sec:bridge}
In bridging the worlds of ethics and law, and considering the algorithmic discrimination of PWDs from both perspectives, we argue that the legal understanding of discrimination is an inherently consequentialist one. Consequentialist morality judges actions as good or bad depending on their consequences only. Intention or character traits and virtues have no power over judgment. In EU law we see a congruent logic, as the intent to discriminate is irrelevant. 
If the effect of an act or practice is to disadvantage a particular group or person, this will amount to discrimination, irrespective of possible well-intentioned or good faith practices. 

From a legal perspective, the consequentialist approach is undisputed, as EU equality law does not aim for any moral verdict of the defendant's character. For ethical deliberations though, consequentialist literature can have a major impact by re-calibrating our moral compass about discrimination against PWDs. For example, arguing with classic utilitarianist John Rawls, if we were to establish a new society tomorrow, the most inclusive and least biased society would be established under the ignorance of one’s own position within this new society \cite{rawls1971theory}. With everyone in mind, including PWDs, people organically agree on principles that guarantee a maximum of individual liberty without restricting the freedom of others (liberty principle) as well as a distribution of equal opportunity to prosper (difference principle). The latter includes a moral obligation to support those who are less advantaged \cite{rawls1971theory}. In a society built on these principles and obligations, we could apply consequentialist jurisdiction as a natural ethical reinforcement to tackle discrimination.

\section{Legal Analysis}\label{sec:legal}
This section examines the relevant applicable legal EU framework and demonstrates how both obstacles and opportunities for social justice stem from its application to AHSs. First, Section \ref{sec:nondiscr} discusses a PWD’s right to \textit{reasonable accommodation} under EU equality law, which entails the possibility for accommodating hiring procedures if needed. The contextual assessment of reasonable accommodation is hardly reconcilable with data-driven AHSs. However, the use of AI in recruiting may have some unexpected benefits for PWDs in terms of proving \textit{prima facie} discrimination, as explained in Section \ref{sec:info_proof}. Second, Section \ref{sec:techlaw} examines the GDPR \cite{GDPR2016} and the AIA \cite{AIA2021}, the latter representing the first attempt by a major regulator to provide a comprehensive framework for AI. We demonstrate that the processing of \textit{disability data} for providing accommodation should be allowed under GDPR. Additionally, we shed light on some new instruments laid down in the AIA that may help fight disability discrimination, such as processing disability data for \textit{debiasing}, and the proposed \textit{public database} for high-risk AI systems. Finally, Section \ref{sec:comp} approaches two recent legislative initiatives in the field of algorithmic discrimination comparatively: the proposed EU Directive on platform work \cite{ECimproves2021} and the New York City bill on automated hiring \cite{NYcitybill}, which aim to provide examples for how algorithmic discrimination should or should not be tackled by the law.

\subsection{EU non-discrimination law}\label{sec:nondiscr}
The risk of algorithmic discrimination in recruitment is globally recognized as a topic of critical importance. Whilst the risk of discrimination for PWDs is widely discussed\footnote{Examples can be found in the debates surrounding the algorithmic profiling system of the Public Employment Service in Austria \cite{allhutter2020algorithmic}, the French public algorithmic student allocation platform \cite{equinet2020}, or the private HireVue hiring platform \cite{sanchez2020does}.}, surprisingly little has been published on the right to reasonable accommodation for PWDs in automated hiring. The difficulties in presenting a claim of direct or indirect discrimination under EU equality law have been addressed in literature in recent years \cite{wachter2021fairness, gerards2021algorithmic}. Whereas evidential requirements are difficult to meet in case of algorithmic discrimination, the need of an effective legal framework is even more pressing in this context where existing inequalities tend to be exacerbated.
 
\subsubsection{Legal sources and scope of application}\label{sec:nondiscr_scope}
Protection against discrimination in Europe is guaranteed by both EU and Council of Europe law. In this paper we mainly focus on the EU \textit{acquis}, most notably Article 21 of the EU Charter of Fundamental Rights, Articles 2 and 10 of the Treaty on the Functioning of the EU, the Employment Equality Directive (2000/78/EC, hereinafter EED) \cite{EED2000}, and jurisprudence of the European Court of Justice (CJEU). Finally, the EU is also a party to the United Nation Convention on the Rights of Persons with Disabilities (CRPD).

Whilst the prohibition of disability discrimination in employment under the EED is well-known, the extension of its scope of application to AHSs warrants further elaboration. The EED also applies to “access to employment [...] including selection criteria and recruitment conditions” (Article 3(a)). This includes advertising job positions and job recommendations \cite{nondiscriminationhb, galinavspeech}, as well as the design architecture of AHSs. Furthermore, the CJEU and the national courts have embraced a rather broad interpretation of Article 3 \cite{nondiscriminationhb}, to the extent that any \textit{de facto} employment relationship should fall within the Directive’s scope, including employment administrations, temporary agency, and platform workers \cite{gerards2021algorithmic}.

Disability is not defined in any EU legal instruments. However, the concept is of utmost importance as the right to reasonable accommodation only applies to PWDs. Nowadays, the spectrum of disability includes situations that had previously remained on its borders. The CJEU has given an autonomous interpretation of the concept of disability, which is in line with the social approach of the CRPD\footnote{Chacón Navas v Eurest Colectividades SA (2006), Case C-13/05; Coleman v Attridge Law (2008), Case C-303/06; Odar v Baxter Deutschland GmbH (2012), Case C-152/11.; Ring v Dansk almennyttigt Boligselskab (2013) DAB, Case C-335/11; Werge v Pro Display A/S, Case C-337/11; Z v A Government Department and the Board of Management of a Community School (2014), Case C-363/12; Glatzel v Cv Freistaat Bayern (2014), Case C-356/12; Kaltoft v Kommunernes Landsforening (2014), Case C-354/13.}, and includes medical or incurable diseases. For that to be the case, two cumulative conditions must be satisfied. Firstly, the illness must result in a physical, mental, intellectual, or sensory impairment which, in interaction with various barriers, is likely to prevent the person concerned from participating fully, effectively and on an equal footing with other workers in working life. Secondly, the illness must be foreseeable or likely to be long-lasting\footnote{Examples of disabilities that have been considered as such by the CJEU or by the Member States include, e.g., auditory impairment, autism, dyslexia, back injuries, HIV, diabetes, severe obesity/obesity, multiple sclerosis, depression, the long-term consequences after cancer treatment.}. This social model is highly context-specific though and thus difficult to embed in automated systems, as explained in Section \ref{sec:technical} by \cite{binns2020apparent}. 

\subsubsection{The refusal of reasonable accommodation in automated hiring amounts to discrimination}\label{sec:nondiscr_acc}
The prohibition of discrimination requires more than a mere refraining from unequal treatment. It may also require an adjustment of an apparently neutral provision or practice if this creates a particular disadvantage for members of a protected group. In particular, the CJEU \cite{defrenne} has acknowledged a transformative function of equality law \cite{muir2018eu} by also requiring private actors to make proactive efforts against discrimination, and which matches our interpretation of inclusion as a society’s moral obligation in Section \ref{sec:moral}. This is particularly evident with regard to disability, since refusing to provide reasonable accommodation constitutes a \textit{sui generis} form of discrimination under Article 5 EED.

The obligation of reasonable accommodation entails a duty to take appropriate measures, if needed, to enable a person with a disability to have access to, participate in, or advance in employment, or to undergo training, unless such measures would impose a disproportionate burden on the employer (Articles 6 EED and 2(4) CRPD). The CRPD definition is interpreted broadly by the CJEU: measures might be material (e.g., special office, installation of a stair lift) as well as immaterial (e.g., reduction of hours \cite{jettering}, exemption from certain tasks \cite{ecvsitaly}). Even though Article 5 EED does not explicitly state that violation of the right to reasonable accommodation constitutes a form of discrimination, this is indeed well established. In fact, refusal of reasonable accommodation is considered a \textit{sui generis} category of discrimination by Article 2(3) CRPD - which takes precedence over the EED and must thus inform its interpretation \cite{jettering} -, the European Court of Human Rights \cite{camvturkey}, and Members States’ legislation and case law \cite{equinet2021}. Accordingly, in the \textit{Ruiz Conejero} case, Advocate General Sharpston reasoned that a dismissal resulting from a failure to respect an accommodation obligation “amount[s] to unlawful discrimination for the purposes of Directive 2000/78”, thus confirming our position \cite{conejero}.

Such obligation also applies to the hiring process, and in a twofold manner. First, the employer is obliged to ensure reasonable accommodation during the application process, e.g., extra time allotted to a person with dyslexia taking a written test. Second, when deciding on a PWD’s application, the employer must assess whether the candidate could perform the job if provided with reasonable accommodation, e.g., by splitting a full-time function into part-time functions for a PWD who cannot work full-time. There are no formal requirements to request reasonable accommodation, which must be provided from the time the PWD requires it. Nonetheless, reasonable accommodation is in principle only offered if the need for accommodation is or could be known, which presupposes disclosure of disability. However, disclosure is challenging from an ethical perspective, as demonstrated in Section \ref{sec:eth_info}. Furthermore, the question of whether, when and to what extent PWDs have to request reasonable accommodation in the context of AHSs remains unanswered, as pointed out in Section \ref{sec:acc}.

A contextual proportionality test is required to ascertain reasonableness of an accommodation. The UN Committee on the Rights of Persons with Disabilities and the CJEU factor in the cost of the proposed adaptation, the organizational impact, influence on the safety of the working environment and the availability of government subsidies \cite{UNcomment}. However, the concept of proportionality is rather vague and highly context-specific due to its dependence on the nature of the company and the job. Also, a wide variety of adjustments is possible. Hence, it is difficult to predict which factors need to be considered and how they would be ranked \cite{sanchez2020does}. As argued in Section \ref{sec:technical}, it must be assumed that AHSs currently do not provide reasonable accommodation for PWDs. Nevertheless, the mere lack of assessing possible reasonable accommodation for a PWD in a recruitment procedure, can amount to a violation of the right to reasonable accommodation and therefore to discrimination.  

In light of the above, a specific application of the duty of reasonable accommodation could be to provide an alternative hiring procedure for PWDs who fear to be treated unequally. In practice, this could be realized by the opportunity to “opt out” of the AI-driven selection procedure, leading to human intervention. However, this raises ethical concerns, since not everyone desires to disclose their disability this way, and “opting out” may paradoxically lead to an unfair advantage as human bias may judge more leniently than its computational counterpart. At the same time, by design AHSs do not take into account this obligation for accommodation, irrespective of a request. The CJEU argued in \textit{Feryn} \cite{feryn} and \textit{Asociaţia Accept} \cite{asociataaccept}, and recently restated in \textit{Rete Lenford} \cite{lgbti}, that actively deterring job applicants belonging to protected groups from applying is to be considered discrimination, even if no recruitment process is on-going. With this, one could argue that discriminatory job recommendation and advertising breach EU equality law as well if they both \textit{in effect} undermine equal access to the labor market \cite{gerards2021algorithmic}. Consequently, applying this framework to the case at hand, AHSs that do not take into account the obligation of reasonable accommodation should be considered \textit{prima facie} discriminatory, regardless of whether accommodation is requested.

\subsubsection{Burden of proof and justification}\label{sec:nondiscr_proof}
The assessment of reasonableness of an accommodation is context-sensitive and open to judicial interpretation. The nature of AI applications and more precisely the fact that output is statistically probable yet inevitably uncertain given the inherent opacity of AI, makes it difficult for claimants to realize they are treated unequally, to gather the necessary information, and to seek redress \cite{binns2020apparent}. Such technical challenges are further developed in Section \ref{sec:providing_acc}. Compared to traditional forms of discrimination, automated discrimination is more abstract and intangible \cite{wachter2021fairness}.

A key element in EU anti-discrimination law is the shifting of the burden of proof, introduced to contrast the difficulty of proving discrimination suffered by protected groups and by virtue of the principle of effectiveness. The claimant must demonstrate facts from which discrimination may be presumed and which the alleged defendant should rebut (Article 10 EED). Traditionally, in discrimination cases, certain elements, e.g., the composition of the comparator group or a particular disadvantage need to be demonstrated in order to meet the threshold to shift the burden of proof. However, these challenging requirements do not apply to reasonable accommodation cases. In these cases, the assumed lack of reasonable accommodation provides sufficient evidence to suggest that discriminatory treatment may have occurred and thus raise a presumption of discrimination. Still, a refusal to make reasonable accommodation can be justified if they result in a disproportionate burden on the employer (Article 5 EED).  

Moreover, the CJEU recognized in 1989 in the landmark \textit{Danfoss} case on equal pay that “where an undertaking applies a system [of pay] which is totally lacking in transparency, it is for the employer to prove that his practice […] is not discriminatory” (par. 16) \cite{danfoss}. Arguably, this principle should be applied to AHSs that lack transparency, so that the principle of equal treatment can be applied effectively to automated hiring processes as well. However, taking into account that a refusal to provide information on reasonable accommodation can raise a presumption of discrimination \cite{galinavspeech}, it might well be that a justification under Article 5 that claims a disproportionate burden will be difficult to provide.

\subsection{EU technology law}\label{sec:techlaw}
Whereas EU non-discrimination law applies to any hiring process, EU technology law also shapes the regulatory environment of AI-driven hiring. For the purposes of this paper, we use “EU technology law” to refer to the GDPR \cite{GDPR2016} and the proposed AIA \cite{AIA2021}. The relevance of the GDPR is evident since AHSs process personal data. Moreover, it does not merely safeguard privacy, it rather fulfills a much broader objective, which is “protecting fundamental rights and freedoms” (Article 2), including the right to non-discrimination (mentioned in Recitals 71 and 75). Given its broad definition of AI systems, all AHSs fall within the scope of application of the AIA (Article 3(1) and Annex I). Moreover, AI-driven recruiting is listed as a high-risk AI in Annex III, and thus is subject to a strict regime. A comprehensive study of the relevant provisions would be outside the scope of this paper. The analysis rather aims to assess whether the legal obligations imposed on employers and AI providers effectively pursue social justice objectives \cite{gabriel2021towards, gabriel2020artificial, selbst2019fairness, wolff2021technology}. On the other hand, technology may inadvertently facilitate the exercise of legal rights, as argued in Section \ref{sec:info_proof}.

\subsubsection{Processing of data related to disability: Problem statement}\label{sec:techlaw_data}
Assuming that one of the conditions laid down in Article 6 GDPR for processing personal data is met, AHSs additionally process disability data. As a general rule, Article 9(1) GDPR classifies health data as sensitive data, and forbids processing. Even though its purpose is to avoid unfair discrimination against data subjects, Section \ref{sec:acc} maintains that “fairness through unawareness” is as simplistic a solution as it is an ineffective one. On the contrary, processing of sensitive data may be necessary for mitigating bias, regardless of how counter-intuitive it might seem \cite{europequalityhb}. Indeed, inequalities are rooted in society \cite{barocas2016big, criado2019digital, dignum2021myth, hildebrandt2021discrimination, lerman2013big, wachter2020bias}, and AI systems merely perpetuate these patterns of injustice \cite{pasquale2015black, o2016weapons, angwin2016machine, eubanks2018automating, crawford2021atlas}, “freezing the future and scaling the past” \cite{hildebrandt2020code}. In particular, as pointed out in Section \ref{sec:aiethics}, PWDs experience specific forms of discrimination in both the analog and the digital world, which also justifies framing the inclusion of PWDs as a society’s moral obligation.

Yet another problem stems from the algorithmic discrimination resulting from proxies (e.g., searching on the internet for a wheelchair, or participating in disability-related volunteering activities), as thoroughly explained in Section \ref{sec:acc}. Such proxies are not personal data in the meaning of the GDPR, and thus fall outside its scope. Nevertheless, the literature has repeatedly pointed out their adverse impact on protected groups \cite{barocas2016big, criado2019digital, hildebrandt2021issue, krollaccountable, nugent2020recruitment, wachter2020bias}. Consequently, having access to the sensitive information stored in proxies might be useful to recognize and mitigate the effect of them. Finally, an employer needs to be aware of someone’s disability to make an informed decision about a reasonable accommodation. Besides it being an ethical challenge riddled with positive and negative consequences, awareness may also be problematic under data protection law, due to the prohibition of the resulting processing of health data which falls under Article 9(1) GDPR.

\subsubsection{Processing of data related to disability: Exceptions}\label{sec:techlaw_exc}
Staying with the GDPR framework, we assess whether it is lawful for employers to process health data regarding job applicants’ disability status, in order to provide them with reasonable accommodation, or to correct an unjustly biased rejection of their application. Article 9(2) provides some exceptions that allow processing of sensitive data. In particular, exceptions a), b), f), and g) are the most relevant for our use case.

The exception based on a data subject’s “explicit consent” under letter a) should not be favored in this case, since consent can hardly be given freely in the hiring context, due to the imbalanced power relationship between the employer and the applicant \cite{WPopinion8, dataprotectionhb}. Rather, processing of sensitive data as “necessary for carrying out the obligations […] in the field of employment” under letter b) is a more suitable option. Arguably, framing reasonable accommodation as a \textit{proactive obligation} for the employer has the implicit consequence of opening the doors to such an exception. Indeed, granting equal opportunities to applicants with a disability is not only an application of the general right to non-discrimination, but also meets the specific duty to provide reasonable accommodation. Of course, this reasoning applies only to discrimination against PWDs, who are the sole beneficiaries of the right to accommodation under EU equality law. As a consequence, employers should be allowed to process an applicant’s sensitive data for evaluating a suitable accommodation.

Nonetheless, exception b) is not available in the event of a \textit{fully automated} decision in the meaning of Article 22 GDPR, which restricts the option to exceptions a) (already excluded) and g). The latter allows processing of sensitive data for reasons of “proportionate substantial public interest”. Even though EU equality law is meant to protect individuals’ rights, it has an undeniable impact on social justice and is thus not only related to but pursues the public interest \cite{mantelero2017group, smuha2021beyond}. Moreover, giving everyone equal opportunities to succeed improves overall productivity, as it gives a chance to highly skilled applicants who might be rejected because of prejudices \cite{wachter2021richer}. However, letter g) requires an explicit authorization by national or Union law. Thus far, the UK seems to be the only European State (though not EU Member) to have implemented such a law \cite{specialcategory, EDPDconsentguideline, binns2021could} that allows processing of sensitive data under 23 conditions for the reason of public interest, including “equality of opportunity or treatment” (Data Protection Act 2018, Schedule 1, par. 8). Therefore, despite being abstractly a good fit for the presented use case in theory, such an exception is \textit{de facto} unavailable right now throughout Europe, except for the UK.

If we consider fully automated AHSs as unlawful in the first place, the issue might lose its relevance and exception b) would become the most pertinent one. Arguably, none of the three criteria allowing fully automated decision-making under Article 22(2) \cite{WPguidelines} is met in this case \cite{sanchez2020does, kim2021artificial, WPopinion8}. If such systems were deployed, candidates could thus allege violation of Article 22(1), regardless of the content of the decision. In a different scenario in which applicants exercise their legal rights under the GDPR or EU equality law, processing of sensitive data is permitted under letter f), allowing it for the purposes of “legal claims”, which should not be intended as limited to court proceedings. 

For what concerns the proposed AIA, Article 10(5) equips providers of high-risk AI systems, such as AHSs, with an additional exception for processing of sensitive data, namely for \textit{debiasing} purposes. This provision may be invoked by the provider of the system, i.e., the same employer or, in most cases, a third party, at a developing stage. Such a broad exception would have both advantages as well as disadvantages \cite{specialcategory}. It may nonetheless respond to criticism raised thus far with regard to the ban on collecting equality data for debiasing purposes \cite{fraopinion, alidadi2017gauging, van2019ethnic, vzliobaite2016using} and hopefully enable AI systems to provide reasonable accommodation as foreseen in Section \ref{sec:info_rec}.

In conclusion, as the law currently stands, processing of data related to applicants’ disabilities should be allowed under the exception laid down in Article 9(2)(b) GDPR. A normative intervention, at an EU or a national level, may be needed for the purposes of the exception under letter g) in the event of fully automated AHSs, even though the very same lawfulness of full automation in recruiting is disputable. Finally, Article 10(5) AIA could also provide a further exception, for processing of sensitive data for debiasing purposes during development stages. 

\subsubsection{Job applicants’ rights}\label{sec:techlaw_app}
Being data subjects for the purposes of the GDPR, job applicants with disabilities have the rights listed in Articles 12-22. In contrast, the AIA does not provide any legal action for individuals affected by an AI system \cite{smuha2021eu}, even though a violation of the prescribed standards may be alleged in a national tort law claim \cite{veale2021demystifying}. These data protection remedies ultimately safeguard data subjects’ fundamental rights (Article 2 GDPR). Therefore, they can be invoked to enforce a PWD's right to non-discrimination. The \textit{right to access processed personal data} (Article 15(1)) might help gather proof for establishing \textit{prima facie} discrimination and thus reverse the burden of proof \cite{hacker2018teaching}, as already mentioned in Section \ref{sec:info_proof}. On the contrary, non-discrimination law does not equip protected groups with such a right to access such data, hence data protection law proves itself a precious instrument for fighting algorithmic discrimination \cite{hacker2018teaching, galinavspeech}.

Additionally, data subjects are granted the \textit{right to receive meaningful information about the logic involved} (Article 15(1)(h)) as well as the \textit{right to contest the decision}, the \textit{right to express one’s point of view}, and the \textit{right to obtain human intervention} (Article 22(3)). Indeed, these provisions could be exercised by PWDs to contest a decision in violation of their right to non-discrimination, as well as to receive information on the functioning of the AHS for discovering whether it might discriminate and thus the opportunity to demand an alternative evaluation procedure. However, the importance of these rights should not be overestimated. Not only is their meaning highly debated in literature \cite{selbst2018meaningful, wachter2017right, malgieri2017right, edwards2018enslaving, mendoza2017right}, but their application is also restricted to fully automated decisions in the meaning of Article 22 \cite{WPguidelines, bygrave2019minding}. Nevertheless, fully automated AHSs would be deemed radically unlawful under Article 22(1) for the above-mentioned reasons.

\subsubsection{Employers’ obligations}\label{sec:techlaw_emp}
Algorithmic disability discrimination constitutes a severe violation of the principles of accuracy and fairness enshrined in Article 5(1) GDPR \cite{hacker2018teaching}. Moreover, Art. 35(1) prescribes that data-controllers, i.e. employers, conduct a \textit{data protection impact assessment} (DPIA) whenever a specific type of processing “is likely to result in a high risk to the rights and freedoms of natural persons”. Briefly, the DPIA is an assessment of those risks and the measures determined to address them \cite{WPguidelinesDPIA}. It is indisputable that AHSs entail a high risk in the meaning of Article 35 and thus require a DPIA. This provision aims to promote further proactive engagement on behalf of the data controller, following a bottom-up approach \cite{dataismine} and ultimately establishing accountability as laid down in Article 5(2), since the specific safeguards are not dictated by law \cite{alemanno2012regulating}. Rather, it is the controller’s duty to implement them. Hence, it can be sustained that both the GDPR and EU disability law aim at promoting a positive effort on behalf of private actors, and ultimately of society \cite{muir2013transformative, wolff2021technology}. If duly conducted, the DPIA provides an analysis of systemic risks of unfairness and discrimination, hence facilitating the exercise of job candidates’ individual rights to contestation, to express their view, and to human intervention \cite{dataismine, quelle2018enhancing, kaminski2020multi}.

Transitioning to the AIA, we are met with the idea of a public, central database for the registration of “standalone” high-risk AI systems, such as AHSs. Such a database could assist advocacy bodies and national authorities in detecting discriminatory AHSs \cite{veale2021demystifying}. Furthermore, being a high-risk AI application, AHSs ought to be “effectively overseen” by a human for “preventing or minimizing the risks to […] fundamental rights” (Article 14). Arguably, this provision would strengthen the employer’s duty to provide reasonable accommodation. In particular, Article 14(4)(d), entrusts the human supervisor with the decision whether to use the high-risk AI system in the first place, or otherwise disregard its output. Therefore, if an AHS posed a risk of algorithmic discrimination against PWDs, the employer would be obliged to provide an alternative evaluation procedure (Articles 5 EED and 14(4)(d) AIA).

\subsection{Comparative insights}\label{sec:comp}
Lawmakers in Europe and in the US are currently occupied with algorithmic discrimination, specifically when occurring in automated hiring. Across the Atlantic, the debate on algorithmic discrimination in hiring has been fostered by the recent enactment of New York City’s bill on automated employment decision tools last December \cite{NYcitybill}. It reveals a transparency-oriented approach to non-discrimination, since it mandates the disclosure of technology used to evaluate an application. However, it seems to fall short of protecting PWDs. In fact, it only requires notice after screening of candidates, hence failing to present applicants with disabilities with the opportunity to ask for accommodation. Evidently, this would only be possible if the PWDs discover \textit{in advance} that the AHS might be unjustly biased against them  \cite{scherer_shetty_2021}. Therefore, the GDPR seems more far-reaching and fair than the New York bill at first glance.

Additionally, some recent developments in the comparable field of \textit{platform work} can provide further insight on the selected use case. Notably, platform workers are subject to algorithms that automatically determine work conditions, i.a., their earnings and sanctions. Following the Spanish \textit{Ley Rider} law No 9/2021 \cite{royaldecreeley}, the EU Commission published its draft EU Directive on platform work last December \cite{ECimproves2021}. Similar to Article 9 GDPR, the proposed Directive prohibits the processing of sensitive data, including the very same exceptions. However, it is interesting to note that the exception based on the worker’s explicit consent (Article 9(2)(a) GDPR) is explicitly excluded here, which confirms its unsuitability as a lawful legal basis for data processing in the employment context \cite{WPopinion8, dataprotectionhb}. Furthermore, the proposal grants platform workers the right to a human review even when the decision is merely supported by an automated tool; this is contrary to Article 22 GDPR that provides such a right only for those subject to fully automated profiling. Such a provision may not only be a source of inspiration for enhancing the protection of PWDs against discrimination, but since AHSs ought not to be fully automated in light of the previous arguments, the right to obtain human intervention should be extended to every and all AHSs for it to be effective.

\section{Technical Analysis}\label{sec:technical}

Besides developments in assistive technology \cite{leo2017computer, mulfari2018tensorflow, stangl2020person, mostofa2020iot, kidzinski2020artificial}, AI is generally developed without the needs of PWDs in mind, giving rise to applications discriminating against them \cite{allen2020artificial, hutchinson2020unintended, trewin2019considerations}. For other protected groups, AI systems can be adjusted to improve the statistical fairness of its output, e.g., by increasing the importance of underrepresented groups \cite{mehrabi2021survey}. Yet such a data-driven solution is hard to implement for PWDs, due to the heterogeneity of disability, as well as the need for a context-specific assessment of its relevance in any given circumstance \cite{guo2020toward, whittaker2019disability}.  

While we acknowledge that technology alone cannot achieve social justice \cite{bennett2020point}, our aim in this section is to assess how AHSs can be made fairer than they currently are for PWDs. First, we propose to reduce the technical obstacles posed by the heterogeneity of disabilities, by using AI to understand disabilities. 
We refer to this process as \textit{modeling disability} in Section \ref{sec:modeling}. Second, in Section \ref{sec:acc} we assert that AI systems should be able to leverage this understanding in order to provide the necessary accommodation and judge its reasonableness. Finally, even if unsuccessful in these tasks, we remain optimistic about the opportunities that AI can offer as a source of information discussed in \ref{sec:info}. 

\subsection{Technical understanding of disability}\label{sec:modeling}
Most of the success of recent AI systems has been due to their capacity to detect patterns in large amounts of data, with which decisions could then be automated at a large scale. Yet, as evidenced by the case where Amazon's hiring system learned to systematically undervalue women \cite{darroch2017amazon}, not all learned patterns are desirable. To prevent discrimination, AI systems are now being modified to avoid patterns involving protected characteristics, again using a data-driven approach \cite{mehrabi2021survey}. Standard solutions in this area attempt to remove bias against a protected group by introducing a reversed, synthetic bias in the form of positive discrimination for the entire group. However, such a debiasing technique can only be successful under the assumption that discrimination is homogeneous: e.g. that women are discriminated "because they are women".

\subsubsection{Heterogeneity}
In contrast to the protected groups that are traditionally considered, the discrimination that PWDs experience is mainly heterogeneous \cite{trewin2018ai}. For instance, an applicant that requires assistive technology to complete an online test may score lower on timing metrics in an AHS \cite{trewin2019considerations}. Comparably, candidates with autism may avoid eye gaze during an interview, which may be misinterpreted as a lack of confidence in their abilities. Finally, someone who has been recovering from a long-term illness in the past may be judged unfairly due to the gap in their résumé. Even these examples are each just one simplified instance of how the specific `disability profile' of a person can lead to their undervaluation in an automated assessment. None of these examples, however, was an AHS discriminatory due to the person having a disability \textit{per se}, but rather because the assessment procedure itself was inadequate for the applicant's specific needs.

Instead, AI systems would do well in giving people the opportunity to make the system aware of their disability in an accessible way. Next, the AI should elaborate on such information to determine whether the specific disability would impact the requested skills and abilities. After all, for the purposes of providing reasonable accommodation (for which we refer to Section \ref{sec:acc}), it is not necessary to know the exact disability profile of an applicant. We should only focus on the specific reasons why they require accommodation in order to be assessed justly. For example, if due to a disability an applicant is unable to maintain eye contact during an interview, it is not relevant whether a visual impairment or autism is the determining factor. In both cases, verbal communication may be more informative about their capabilities.

\subsubsection{Modeling disability}
As a possible technical solution for managing the heterogeneity of disability, we therefore urge for more research into \textit{modeling disability}, i.e. AI systems that can extract a set of features from an individual disability profile, which we refer to as \textit{limitations} and we assume to be transferable to a wide range of cases. The idea is that a different AI (sub)system can rely on such information to propose reasonable accommodation while being unaware of the individual's sensitive disability profile. Examples of such limitations may include "being wheelchair-bound", "unable to communicate verbally" or "unable to work in loud environments". 

Inspiration may be found in recent advances in AI-based human resources (HR) tools that aim to model the skills and competencies of users \cite{gugnani2020implicit}. A well-known example is the use of such a framework by LinkedIn to more efficiently profile job seekers \cite{ha2015personalized, ramanath2018towards}. The main idea of this framework is to construct an ontology of skills, of which some are then assigned to people based on their own reporting or on automatic extraction from their résumés. Unfortunately, little is known about how these frameworks currently operate due to privacy concerns and economic motivations \cite{gugnani2020implicit}. Still, it may be possible to better understand the disability of a person by being able to map it, manually or through automation, to an ontology of limitations. Arguably, the technical challenge is in constructing this ontology with the right level of granularity while taking into account that some disabilities only posed limitations in the past, e.g., an illness from which a person has fully recovered. 

At the same time, we ought to be aware of the risks in attempting to model disabilities. Though it may help in simplifying the automation of a hiring process, the same simplification may lead to a characterization of limitations that is too vague to provide a true understanding of the disability, or one that is too rigid to provide the nuance that a PWD deserves. Fundamentally, the standardization of a person's abilities poses ethical concerns (already discussed in Section \ref{sec:aiethics}) that should be carefully balanced against the potential practical advantage that such standardization provides \cite{whittaker2019disability}. A system to model disabilities should thus be careful to always put the holistic needs of the individual first. 

\subsection{Reasonable accommodation}\label{sec:acc}
Given the myriad of ways in which data and machine learning models can be biased \cite{mehrabi2021survey}, we should assume that AHSs will not fairly evaluate PWDs unless the process is specifically designed to reduce discrimination. A well-known result from fairness research is that simply removing sensitive data from the algorithm's input (achieving the so-called "fairness through unawareness") is insufficient, since the sensitive information relating to an individual may still be correlated with other, non-protected, features, or even proxies, in a subtle manner that ultimately cannot be ignored \cite{pedreshi2008discrimination}. Similar to Section \ref{sec:modeling}, we are concerned with the specific heterogeneous risks that PWDs face during these processes. We will explore several types of accommodations and then evaluate their reasonableness.

\subsubsection{Providing accommodation}\label{sec:providing_acc}
Under the assumption that the AI system has a practical model of a person's disability as discussed in Section \ref{sec:modeling}, we distinguish four ways in which a standard AHS may need to be modified in order to tackle the heterogeneity of disabilities and provide reasonable accommodation. First, it may be necessary to adjust the predictions of a model due to a group of persons with a particular disability having been undervalued in the past. Second, some disabilities may render the standard hiring process simply inaccessible, e.g., mutism if verbal communication is part of the assessment. Third, failing to automate accommodation, the only solution may be replacing the AHS with a human. Fourth, even if the candidate is fairly assessed, they may still require accommodation in the future to perform their job.

\paragraph{Adjusting predictions}
The adjustment of predictions is the closest to the standard fairness modifications made for other protected groups. It is necessary when the data or model are imperfect \cite{mehrabi2021survey}. For example, a person that suffered from a long-term illness may have only been employed part-time in the past. Especially if they have now recovered, they should not only be recommended part-time jobs in the future. In a different example, we may see that a data-driven AI system fails to provide any sensible assessment for people who use a wheelchair, simply because the model is unable to learn which jobs are accessible to those individuals and which are not. The fact that the relevance of limitations is job-specific makes a data-driven approach particularly difficult.

\paragraph{Multimodal input}
A hiring process may be setup in such a way that it, in theory, gathers enough information about a person without job-related limitations in order to make a fair assessment. However, if it involves a verbal interview for example, a person needing sign language interpretation can never be fairly judged based on audio data alone. It is thus necessary to design the hiring procedure in such a way that accepts input from multiple sources in different formats, e.g., evaluating either audio or sign language video data. Accommodation is then provided for some PWDs by excusing them from being tested in a format that is inaccessible to them. The problem where people need to receive the same assessment, even though they are `seen' in different types of input format, is referred to as a \textit{multimodal} problem \cite{baltruvsaitis2018multimodal}. While perceiving the world in this way is natural for humans, it is not the norm for AI systems and is therefore also not readily accessible for employers to use.

\paragraph{Human intervention}
An AI system should have the ability to judge whether it can grant adequate accommodation. Otherwise, a human should replace it. However, this is infeasible for platforms where decisions are automated on a massive scale like LinkedIn \cite{kenthapadi2017personalized}. Moreover, it may not improve fairness for PWDs, because there is no guarantee that the human will provide a less discriminatory or biased assessment. Arguably, the main advantage is a legal one: as discussed in Section \ref{sec:techlaw}, automated decisions are held to a higher standard than those taken by humans.

\paragraph{Future accommodation}
We finally mention the problem of providing accommodation for the actual job. A classic example is a person who is wheelchair-bound and thus may require changes to the architecture at the work place to properly perform their job. In order to also automate the end of the hiring process, AI should be able to make recommendations so that the job is made reasonably accessible.

\subsubsection{Estimating reasonableness of accommodations}
For every type of accommodation to the hiring process listed in Section \ref{sec:providing_acc}, we can question whether it is reasonable to make such adjustments. There is no established legal standard as to what is reasonable, since every accommodation should ideally be judged on a case-by-case basis. 
To judge the reasonableness of accommodations at the workplace, AI systems could be used to predict its economic cost for the employer. If this cost is accurately estimated as unreasonable, then the result could be used by the employer as motivation to not grant that accommodation. Building and training a prediction algorithm is again challenging due to the heterogeneity of disability and of possible tasks and assignments. Additionally, the reasonableness of such accommodations also changes over time as technology improves. 
It should also be questioned which adjustments to the AHS itself are reasonable. The adjustment of predictions for every kind of limitation or the possibility of learning from multimodal input are open research problems. Even if possible with the state-of-the-art, such accommodation carries with it a significant added complexity of the AI system and additional data requirements.




\subsection{AI as a source of information}\label{sec:info}
So far, our technical analysis has mainly consisted of listing the challenges and risks in providing reasonable accommodation in AHSs. However, some of the accommodations we deem ethically and legally necessary may not yet be possible to automate with present technology, due to the heterogeneity of disability that is so incongruent with the current paradigm of machine learning that is consistently data-driven.

Though we hope for further research into ways to model disabilities and provide reasonable accommodation, we also point out two ways in which AI systems can already improve the experience of PWDs in the job market. First, we can leverage data-driven approaches to recommend reasonable accommodations to job seekers, employees and employers. Second, we can benefit from the (partial) automation of hiring processes to protect the rights of PWDs.

\subsubsection{Recommending accommodation}\label{sec:info_rec}
Most hiring tools today are not fully automated (and should not be, as argued in Section \ref{sec:techlaw_exc}). Instead, AI-based tools for hiring are developed to replace specific intermediate steps in the end-to-end hiring process \cite{bogen2018help}. In a semi-automated hiring process where humans still make the most important decisions, AI tools need not provide reasonable accommodation directly. 

Rather, if a candidate chooses to disclose their disability, an AI-based tool could signal the need for accommodation to a human decision maker. If the disability is understood well enough by the AI system (e.g., using a technique discussed in Section \ref{sec:modeling}), then the AI can only report the relevant limitations while keeping the sensitive disability information hidden from the human decision maker. If they agree that accommodation is needed, AI systems could then be used to recommend accommodations. Of the examples listed in Section \ref{sec:acc}, the simplest accommodation to suggest is that of \textit{future} accommodation, as it does not require any modification to the assessment procedure. As discussed in Section \ref{sec:eth_info}, prospective employees also benefit from such recommendations, as information can help establish trust between them and their employers. AI can provide an invaluable benefit to both parties by making the information on reasonable accommodations more easily accessible.

\subsubsection{Protecting rights}\label{sec:info_proof}
If a part of the hiring process is automated through the use of a sufficiently transparent AI, then the use of it may inadvertently facilitate proving discrimination. While gathering proof is arguably one of the most burdensome tasks for PWDs under EU non-discrimination law, in the future it may be gathered from the \textit{ex post} analysis of the automated decision-making process. That way it can, e.g., be established whether individuals with a certain limitation are consistently and unjustly undervalued. As data subjects under EU technology law (see Section \ref{sec:techlaw}), PWDs are also facilitated in exercising their right to access and right to contest. As such, the AI itself may be the surprise witness of an alleged discrimination. 

\section{Conclusion}
Fundamentally, PWDs form a distinguishable vulnerable group that experience separation and discrimination in a manner unlike other protected minorities. As job seekers, PWDs are uniquely disadvantaged in that the disclosure of their disability is difficult to control and may often not be in their best interest, causing a lack of trust in prospective employers. Conversely, these employers ought to feel morally obligated to prevent algorithmic discrimination if they employ AHSs, even though their good faith does not excuse breaches of EU equality law by virtue of its consequentialist approach. 

The attention of lawmakers directed towards algorithmic discrimination in general, and to AHSs specifically, further demonstrates the urgency of the matter. Our analysis of current EU non-discrimination law concludes that the right of PWDs to reasonable accommodation also applies to cases in which AHSs are used. EU technology law provides instruments to ensure reasonable accommodation in automated hiring, paving the way to utilizing disability data for the purposes of improving social justice, and imposing further obligations on employers that may help enforce the applicants’ right to non-discrimination. Though there might be no determined legal solution to algorithmic discrimination, the proposed AIA (and the proposed Directive on platform work) reveal(s) the EU Commission’s willingness to address algorithmic bias.

In practice, AI systems struggle to supply reasonable accommodation due to a heterogeneity that is unique to PWDs as a protected group, which makes a data-driven solution infeasible. To some extent, this heterogeneity can be tackled by employing AI to better understand disabilities. Yet even with this understanding, the technical adjustments needed to provide accommodation are varied and challenging, presenting obvious risks for PWDs and therefore a demand for further research and testing in the long term. In the short term, AI can offer opportunities for PWDs by acting as a source of information during the hiring process, either by assisting in providing reasonable accommodation in hiring decisions that are not automated, or by presenting proof of the lack of accommodation in hiring decisions that are automated.

Our analysis has shed some much-needed light on the many forms in which PWDs experience algorithmic discrimination by AHSs. The use of data-driven tools for hiring purposes deserves careful consideration, given their impact on an individual’s life. Decent work and employment are vital for the well-being and dignity of all.

\begin{acks}
We thank Mark Dempsey for his fruitful comments. This research was supported by the Flemish Government (AI Research Program), the BOF of Ghent University (PhD scholarship BOF20/DOC/144) and Roma Tre University.
\end{acks}

\bibliographystyle{ACM-Reference-Format}
\bibliography{references}


\begin{thebibliography}{108}


\ifx \showCODEN    \undefined \def \showCODEN     #1{\unskip}     \fi
\ifx \showDOI      \undefined \def \showDOI       #1{#1}\fi
\ifx \showISBNx    \undefined \def \showISBNx     #1{\unskip}     \fi
\ifx \showISBNxiii \undefined \def \showISBNxiii  #1{\unskip}     \fi
\ifx \showISSN     \undefined \def \showISSN      #1{\unskip}     \fi
\ifx \showLCCN     \undefined \def \showLCCN      #1{\unskip}     \fi
\ifx \shownote     \undefined \def \shownote      #1{#1}          \fi
\ifx \showarticletitle \undefined \def \showarticletitle #1{#1}   \fi
\ifx \showURL      \undefined \def \showURL       {\relax}        \fi
\providecommand\bibfield[2]{#2}
\providecommand\bibinfo[2]{#2}
\providecommand\natexlab[1]{#1}
\providecommand\showeprint[2][]{arXiv:#2}

\bibitem[\protect\citeauthoryear{??}{cam}{2016}]%
        {camvturkey}
 \bibinfo{year}{51500/08, \c{C}am v Turkey, European Court of Human Rights,
  Judgement of 23 February 2016}\natexlab{}.
\newblock \bibinfo{howpublished}{[Online]}.
\newblock
\newblock
\shownote{Available:
  \url{https://www.legislationline.org/download/id/7067/file/ECHR_Cam_v_Turkey_2016_en.pdf}.}


\bibitem[\protect\citeauthoryear{??}{dan}{1989}]%
        {danfoss}
 \bibinfo{year}{C-109/88, Handels- og Kontorfunktionærernes Forbund I Danmark
  v Danfoss represented by Dansk Arbejdsgiverforening, Judgement of 17 October
  1989}\natexlab{}.
\newblock \bibinfo{howpublished}{[Online]}.
\newblock
\newblock
\shownote{Available:
  \url{https://curia.europa.eu/juris/liste.jsf?language=de&num=C-109/88&td=ALL}.}


\bibitem[\protect\citeauthoryear{??}{con}{2018}]%
        {conejero}
 \bibinfo{year}{C-270/16, Carlos Enrique Ruiz Conejero v Ferroser Servicios
  Auxiliares SA and Ministerio Fiscal, Judgement of 18 January
  2018}\natexlab{}.
\newblock \bibinfo{howpublished}{[Online]}.
\newblock
\newblock
\shownote{Available:
  \url{https://curia.europa.eu/juris/document/document.jsf?text=&docid=198527}.}


\bibitem[\protect\citeauthoryear{??}{ecv}{2013}]%
        {ecvsitaly}
 \bibinfo{year}{C-312/11, European Commission v Republic of Italy, Judgement of
  4 July 2013}\natexlab{}.
\newblock \bibinfo{howpublished}{[Online]}.
\newblock
\newblock
\shownote{Available:
  \url{https://curia.europa.eu/juris/document/document.jsf?text=&docid=139105}.}


\bibitem[\protect\citeauthoryear{??}{jet}{2013}]%
        {jettering}
 \bibinfo{year}{C-335/11 and C-337/11, Jette Ring, §29-30, Judgement of 11
  April 2013}\natexlab{}.
\newblock \bibinfo{howpublished}{[Online]}.
\newblock
\newblock
\shownote{Available:
  \url{https://curia.europa.eu/juris/document/document.jsf?docid=136161&doclang=EN}.}


\bibitem[\protect\citeauthoryear{??}{gal}{2012}]%
        {galinavspeech}
 \bibinfo{year}{C-415/10, Galina Meister v Speech Design Carrier Systems GmbH,
  Judgement of 19 April 2012}\natexlab{}.
\newblock \bibinfo{howpublished}{[Online]}.
\newblock
\newblock
\shownote{Available:
  \url{https://eur-lex.europa.eu/legal-content/EN/TXT/?qid=1490004730485&uri=CELEX:62010CJ0415}.}


\bibitem[\protect\citeauthoryear{??}{def}{1976}]%
        {defrenne}
 \bibinfo{year}{C-43/75, Gabrielle Defrenne v Soci\'{e}t\'{e} anonyme belge de
  navigation a\'{e}rienne Sabena, Judgement of 8 April 1976}\natexlab{}.
\newblock \bibinfo{howpublished}{[Online]}.
\newblock
\newblock
\shownote{Available:
  \url{https://curia.europa.eu/juris/liste.jsf?language=de&jur=C,T,F&num=C-43/75}.}


\bibitem[\protect\citeauthoryear{??}{lgb}{2020}]%
        {lgbti}
 \bibinfo{year}{C-507/18, Associazione Avvocatura per i diritti LGBTI v Rete
  Lenford, Corte suprema di cassazione, Judgement of 23 April 2020}\natexlab{}.
\newblock \bibinfo{howpublished}{[Online]}.
\newblock
\newblock
\shownote{Available:
  \url{https://eur-lex.europa.eu/legal-content/EN/TXT/HTML/?uri=ecli:ECLI\%3AEU\%3AC\%3A2020\%3A289}.}


\bibitem[\protect\citeauthoryear{??}{fer}{2008}]%
        {feryn}
 \bibinfo{year}{C-54/07, Centrum voor gelijkheid van kansen en voor
  racismebestrijding v Feryn NV, Arbeidshof te Brussel, Judgement of 10 July
  2008}\natexlab{}.
\newblock \bibinfo{howpublished}{[Online]}.
\newblock
\newblock
\shownote{Available:
  \url{https://eur-lex.europa.eu/legal-content/DE/TXT/?uri=CELEX\%3A62007CJ0054}.}


\bibitem[\protect\citeauthoryear{??}{aso}{2013}]%
        {asociataaccept}
 \bibinfo{year}{C-81/12, Asocia\c{t}ia Accept v Consiliul Na\c{t}ional pentru
  Combaterea Discrimin\u{a}rii, Curte de Apel Bucure\c{s}ti, Judgement of 25
  April 2013}\natexlab{}.
\newblock \bibinfo{howpublished}{[Online]}.
\newblock
\newblock
\shownote{Available:
  \url{https://eur-lex.europa.eu/legal-content/DE/TXT/?uri=CELEX\%3A62012CJ0081}.}


\bibitem[\protect\citeauthoryear{??}{roy}{2021}]%
        {royaldecreeley}
 \bibinfo{year}{Real Decreto-ley 9/2021, Modificaci\'{o}n de la Ley del
  Estatuto de los Trabajadores, Judgement of 11 May 2021}\natexlab{}.
\newblock \bibinfo{howpublished}{[Online]}.
\newblock
\newblock
\shownote{Available: \url{https://www.boe.es/eli/es/rdl/2021/05/11/9}.}


\bibitem[\protect\citeauthoryear{Alemanno}{Alemanno}{2012}]%
        {alemanno2012regulating}
\bibfield{author}{\bibinfo{person}{Alberto Alemanno}.}
  \bibinfo{year}{2012}\natexlab{}.
\newblock \showarticletitle{Regulating the European Risk Society}.
\newblock In \bibinfo{booktitle}{\emph{Better Business Regulation in a Risk
  Society}}. \bibinfo{publisher}{Springer}, \bibinfo{pages}{37--56}.
\newblock


\bibitem[\protect\citeauthoryear{Alidadi}{Alidadi}{2017}]%
        {alidadi2017gauging}
\bibfield{author}{\bibinfo{person}{Katayoun Alidadi}.}
  \bibinfo{year}{2017}\natexlab{}.
\newblock \showarticletitle{Gauging process towards equality? Challenges and
  best practices of equality data collection in the EU}.
\newblock \bibinfo{journal}{\emph{Eur Equality Law Rev}}
  \bibinfo{volume}{2017}, \bibinfo{number}{2} (\bibinfo{year}{2017}),
  \bibinfo{pages}{15--27}.
\newblock


\bibitem[\protect\citeauthoryear{Allen and Masters}{Allen and Masters}{2020}]%
        {allen2020artificial}
\bibfield{author}{\bibinfo{person}{Robin Allen} {and} \bibinfo{person}{Dee
  Masters}.} \bibinfo{year}{2020}\natexlab{}.
\newblock \showarticletitle{Artificial Intelligence: the right to protection
  from discrimination caused by algorithms, machine learning and automated
  decision-making}. In \bibinfo{booktitle}{\emph{ERA Forum}},
  Vol.~\bibinfo{volume}{20}. Springer, \bibinfo{pages}{585--598}.
\newblock


\bibitem[\protect\citeauthoryear{Allhutter, Cech, Fischer, Grill, and
  Mager}{Allhutter et~al\mbox{.}}{2020}]%
        {allhutter2020algorithmic}
\bibfield{author}{\bibinfo{person}{Doris Allhutter}, \bibinfo{person}{Florian
  Cech}, \bibinfo{person}{Fabian Fischer}, \bibinfo{person}{Gabriel Grill},
  {and} \bibinfo{person}{Astrid Mager}.} \bibinfo{year}{2020}\natexlab{}.
\newblock \showarticletitle{Algorithmic profiling of job seekers in Austria:
  how austerity politics are made effective}.
\newblock \bibinfo{journal}{\emph{Frontiers in big Data}}  \bibinfo{volume}{3}
  (\bibinfo{year}{2020}), \bibinfo{pages}{5}.
\newblock


\bibitem[\protect\citeauthoryear{Angwin, Larson, Mattu, and Kirchner}{Angwin
  et~al\mbox{.}}{2016}]%
        {angwin2016machine}
\bibfield{author}{\bibinfo{person}{Julia Angwin}, \bibinfo{person}{Jeff
  Larson}, \bibinfo{person}{Surya Mattu}, {and} \bibinfo{person}{Lauren
  Kirchner}.} \bibinfo{year}{2016}\natexlab{}.
\newblock \showarticletitle{Machine bias: There’s software used across the
  country to predict future criminals}.
\newblock \bibinfo{journal}{\emph{And it’s biased against blacks.
  ProPublica}}  \bibinfo{volume}{23} (\bibinfo{year}{2016}),
  \bibinfo{pages}{77--91}.
\newblock


\bibitem[\protect\citeauthoryear{Aran, van Nuenen, Such, Cot{\'e}, and
  Pacheco}{Aran et~al\mbox{.}}{2020}]%
        {aran2020bias}
\bibfield{author}{\bibinfo{person}{Xavier~Ferrer Aran}, \bibinfo{person}{Tom
  van Nuenen}, \bibinfo{person}{Jose Such}, \bibinfo{person}{Mark Cot{\'e}},
  {and} \bibinfo{person}{Natalia~Criado Pacheco}.}
  \bibinfo{year}{2020}\natexlab{}.
\newblock \showarticletitle{Bias and Discrimination in AI: a cross-disciplinary
  perspective}.
\newblock \bibinfo{journal}{\emph{IEEE Technology and Society Magazine}}
  \bibinfo{volume}{40}, \bibinfo{number}{2} (\bibinfo{year}{2020}).
\newblock


\bibitem[\protect\citeauthoryear{Baltru{\v{s}}aitis, Ahuja, and
  Morency}{Baltru{\v{s}}aitis et~al\mbox{.}}{2018}]%
        {baltruvsaitis2018multimodal}
\bibfield{author}{\bibinfo{person}{Tadas Baltru{\v{s}}aitis},
  \bibinfo{person}{Chaitanya Ahuja}, {and} \bibinfo{person}{Louis-Philippe
  Morency}.} \bibinfo{year}{2018}\natexlab{}.
\newblock \showarticletitle{Multimodal machine learning: A survey and
  taxonomy}.
\newblock \bibinfo{journal}{\emph{IEEE transactions on pattern analysis and
  machine intelligence}} \bibinfo{volume}{41}, \bibinfo{number}{2}
  (\bibinfo{year}{2018}), \bibinfo{pages}{423--443}.
\newblock


\bibitem[\protect\citeauthoryear{Bariffi}{Bariffi}{2021}]%
        {bariffi2021artificial}
\bibfield{author}{\bibinfo{person}{Francisco~Jose Bariffi}.}
  \bibinfo{year}{2021}\natexlab{}.
\newblock \showarticletitle{Artificial Intelligence, Human Rights and
  Disability}.
\newblock \bibinfo{journal}{\emph{Pensar-Revista de Ci{\^e}ncias
  Jur{\'\i}dicas}} \bibinfo{volume}{26}, \bibinfo{number}{2}
  (\bibinfo{year}{2021}).
\newblock


\bibitem[\protect\citeauthoryear{Barocas and Selbst}{Barocas and
  Selbst}{2016}]%
        {barocas2016big}
\bibfield{author}{\bibinfo{person}{Solon Barocas} {and}
  \bibinfo{person}{Andrew~D Selbst}.} \bibinfo{year}{2016}\natexlab{}.
\newblock \showarticletitle{Big data's disparate impact}.
\newblock \bibinfo{journal}{\emph{Calif. L. Rev.}}  \bibinfo{volume}{104}
  (\bibinfo{year}{2016}), \bibinfo{pages}{671}.
\newblock


\bibitem[\protect\citeauthoryear{Bennett and Keyes}{Bennett and Keyes}{2020}]%
        {bennett2020point}
\bibfield{author}{\bibinfo{person}{Cynthia~L Bennett} {and} \bibinfo{person}{Os
  Keyes}.} \bibinfo{year}{2020}\natexlab{}.
\newblock \showarticletitle{What is the point of fairness? Disability, AI and
  the complexity of justice}.
\newblock \bibinfo{journal}{\emph{ACM SIGACCESS Accessibility and Computing}}
  \bibinfo{number}{125} (\bibinfo{year}{2020}), \bibinfo{pages}{1--1}.
\newblock


\bibitem[\protect\citeauthoryear{Binns}{Binns}{2020}]%
        {binns2020apparent}
\bibfield{author}{\bibinfo{person}{Reuben Binns}.}
  \bibinfo{year}{2020}\natexlab{}.
\newblock \showarticletitle{On the apparent conflict between individual and
  group fairness}. In \bibinfo{booktitle}{\emph{Proceedings of the 2020
  conference on fairness, accountability, and transparency}}.
  \bibinfo{pages}{514--524}.
\newblock


\bibitem[\protect\citeauthoryear{Binns and Kirkham}{Binns and Kirkham}{2021}]%
        {binns2021could}
\bibfield{author}{\bibinfo{person}{Reuben Binns} {and} \bibinfo{person}{Reuben
  Kirkham}.} \bibinfo{year}{2021}\natexlab{}.
\newblock \showarticletitle{How Could Equality and Data Protection Law Shape AI
  Fairness for People with Disabilities?}
\newblock \bibinfo{journal}{\emph{ACM Transactions on Accessible Computing
  (TACCESS)}} \bibinfo{volume}{14}, \bibinfo{number}{3} (\bibinfo{year}{2021}),
  \bibinfo{pages}{1--32}.
\newblock


\bibitem[\protect\citeauthoryear{Board}{Board}{2020}]%
        {EDPDconsentguideline}
\bibfield{author}{\bibinfo{person}{European Data~Protection Board}.}
  \bibinfo{year}{2020}\natexlab{}.
\newblock \bibinfo{title}{Guidelines 05/2020 on consent under Regulation
  2016/679}.
\newblock \bibinfo{howpublished}{[Online]}.
\newblock
\urldef\tempurl%
\url{https://edpb.europa.eu/sites/default/files/files/file1/edpb_guidelines_202005_consent_en.pdf}
\showURL{%
\tempurl}


\bibitem[\protect\citeauthoryear{Bogen and Rieke}{Bogen and Rieke}{2018}]%
        {bogen2018help}
\bibfield{author}{\bibinfo{person}{Miranda Bogen} {and} \bibinfo{person}{Aaron
  Rieke}.} \bibinfo{year}{2018}\natexlab{}.
\newblock \showarticletitle{Help wanted: An exploration of hiring algorithms,
  equity, and bias}.
\newblock
  \bibinfo{howpublished}{\url{https://apo.org.au/sites/default/files/resource-files/2018-12/apo-nid210071.pdf}}.
\newblock  (\bibinfo{year}{2018}).
\newblock


\bibitem[\protect\citeauthoryear{Bygrave}{Bygrave}{2019}]%
        {bygrave2019minding}
\bibfield{author}{\bibinfo{person}{Lee~A Bygrave}.}
  \bibinfo{year}{2019}\natexlab{}.
\newblock \showarticletitle{Minding the Machine v2. 0: The EU General Data
  Protection Regulation and Automated Decision-Making}.
\newblock In \bibinfo{booktitle}{\emph{Algorithmic Regulation}}.
  \bibinfo{publisher}{Oxford University Press}, \bibinfo{pages}{248--262}.
\newblock


\bibitem[\protect\citeauthoryear{Commission}{Commission}{2021a}]%
        {ECimproves2021}
\bibfield{author}{\bibinfo{person}{European Commission}.}
  \bibinfo{year}{2021}\natexlab{a}.
\newblock \bibinfo{title}{Commission proposals to improve the working
  conditions of people working through digital labour platforms}.
\newblock
\newblock
\urldef\tempurl%
\url{https://ec.europa.eu/commission/presscorner/detail/en/ip_21_6605}
\showURL{%
\tempurl}


\bibitem[\protect\citeauthoryear{Commission}{Commission}{2021b}]%
        {AIA2021}
\bibfield{author}{\bibinfo{person}{European Commission}.}
  \bibinfo{year}{2021}\natexlab{b}.
\newblock \showarticletitle{Proposal for a Regulation laying down harmonised
  rules on Artificial Intelligence (Artificial Intelligence Act) and amending
  certain Union legislative acts}.
\newblock \bibinfo{journal}{\emph{COM(2021) 206 final}} (\bibinfo{year}{2021}).
\newblock
\urldef\tempurl%
\url{https://eur-lex.europa.eu/legal-content/EN/TXT/?uri=CELEX\%3A52021PC0206}
\showURL{%
\tempurl}


\bibitem[\protect\citeauthoryear{Council}{Council}{line}]%
        {NYcitybill}
\bibfield{author}{\bibinfo{person}{New York~City Council}.} \bibinfo{year}{2021
  [Online]}\natexlab{}.
\newblock \bibinfo{title}{A Local Law to amend the administrative code of the
  city of New York, in relation to automated employment decision tools,
  2021/144}.
\newblock
\newblock
\urldef\tempurl%
\url{https://legistar.council.nyc.gov/LegislationDetail.aspx?ID=4344524&GUID=B051915D-A9AC-451E-81F8-6596032FA3F9}
\showURL{%
\tempurl}


\bibitem[\protect\citeauthoryear{Council~of Europe}{Council~of Europe}{2018}]%
        {dataprotectionhb}
\bibfield{author}{\bibinfo{person}{European Data Protection Supervisor European
  Union Agency for Fundamental~Rights Council~of Europe, European Court of
  Human~Rights}.} \bibinfo{year}{2018}\natexlab{}.
\newblock \bibinfo{booktitle}{\emph{Handbook on European data protection law}}.
\newblock \bibinfo{publisher}{Publications Office of the European Union},
  \bibinfo{address}{Luxembourg}.
\newblock


\bibitem[\protect\citeauthoryear{Crawford}{Crawford}{2021}]%
        {crawford2021atlas}
\bibfield{author}{\bibinfo{person}{Kate Crawford}.}
  \bibinfo{year}{2021}\natexlab{}.
\newblock \bibinfo{booktitle}{\emph{The Atlas of AI}}.
\newblock \bibinfo{publisher}{Yale University Press}.
\newblock


\bibitem[\protect\citeauthoryear{Criado and Such}{Criado and Such}{2019}]%
        {criado2019digital}
\bibfield{author}{\bibinfo{person}{Natalia Criado} {and}
  \bibinfo{person}{Jose~M Such}.} \bibinfo{year}{2019}\natexlab{}.
\newblock \showarticletitle{Digital discrimination}.
\newblock In \bibinfo{booktitle}{\emph{Algorithmic Regulation}}.
  \bibinfo{publisher}{OUP}.
\newblock


\bibitem[\protect\citeauthoryear{Dastin}{Dastin}{2018}]%
        {darroch2017amazon}
\bibfield{author}{\bibinfo{person}{Jeffrey Dastin}.}
  \bibinfo{year}{2018}\natexlab{}.
\newblock \showarticletitle{Amazon scraps secret AI recruiting tool that showed
  bias against women}.
\newblock \bibinfo{journal}{\emph{Reuters}} (\bibinfo{date}{Oct.}
  \bibinfo{year}{2018}).
\newblock
\urldef\tempurl%
\url{https://www.reuters.com/article/us-amazon-com-jobs-automation-insight-idUSKCN1MK08G}
\showURL{%
\tempurl}


\bibitem[\protect\citeauthoryear{De~Gregorio and Dunn}{De~Gregorio and
  Dunn}{2021}]%
        {dataismine}
\bibfield{author}{\bibinfo{person}{Giovanni De~Gregorio} {and}
  \bibinfo{person}{Pietro Dunn}.} \bibinfo{year}{2021}\natexlab{}.
\newblock \bibinfo{title}{Profiling under Risk-based Regulation: Bringing
  together the GDPR and the DSA}.  (\bibinfo{year}{2021}).
\newblock
\urldef\tempurl%
\url{https://assets.ctfassets.net/iapmw8ie3ije/5EuxLPaUIsgGt7R6PgeuFK/c9269e55e10bb2a7a0b392624c08f4d0/De_Gregorio_Dunn_My_Data_is_Mine__1_.pdf}
\showURL{%
\tempurl}
\newblock
\shownote{Unpublished paper.}


\bibitem[\protect\citeauthoryear{Dignum}{Dignum}{2021}]%
        {dignum2021myth}
\bibfield{author}{\bibinfo{person}{Virginia Dignum}.}
  \bibinfo{year}{2021}\natexlab{}.
\newblock \showarticletitle{The Myth of Complete AI-Fairness}. In
  \bibinfo{booktitle}{\emph{International Conference on Artificial Intelligence
  in Medicine}}. Springer, \bibinfo{pages}{3--8}.
\newblock


\bibitem[\protect\citeauthoryear{Dixon, Clancy, Miller, Hoegberg, Lewis, and
  {B. Bender et al.}}{Dixon et~al\mbox{.}}{line}]%
        {equinet2021}
\bibfield{author}{\bibinfo{person}{L. Dixon}, \bibinfo{person}{N. Clancy},
  \bibinfo{person}{B.~M. Miller}, \bibinfo{person}{S. Hoegberg},
  \bibinfo{person}{M.~M. Lewis}, {and} \bibinfo{person}{{B. Bender et al.}}}
  \bibinfo{year}{2021 [Online]}\natexlab{}.
\newblock \bibinfo{title}{Reasonable Accommodation for Persons with
  Disabilities: Exploring Challenges Concerning its Practical Implementation}.
\newblock \bibinfo{howpublished}{An Equinet Publication}.
\newblock
\urldef\tempurl%
\url{https://equineteurope.org/wp-content/uploads/2021/03/Reasonable-Accommodation-Disability-Discussion-Paper.pdf}
\showURL{%
\tempurl}


\bibitem[\protect\citeauthoryear{Edwards and Veale}{Edwards and Veale}{2018}]%
        {edwards2018enslaving}
\bibfield{author}{\bibinfo{person}{Lilian Edwards} {and}
  \bibinfo{person}{Michael Veale}.} \bibinfo{year}{2018}\natexlab{}.
\newblock \showarticletitle{Enslaving the algorithm: From a “Right to an
  Explanation” to a “Right to Better Decisions”?}
\newblock \bibinfo{journal}{\emph{IEEE Security \& Privacy}}
  \bibinfo{volume}{16}, \bibinfo{number}{3} (\bibinfo{year}{2018}),
  \bibinfo{pages}{46--54}.
\newblock


\bibitem[\protect\citeauthoryear{Eubanks}{Eubanks}{2018}]%
        {eubanks2018automating}
\bibfield{author}{\bibinfo{person}{Virginia Eubanks}.}
  \bibinfo{year}{2018}\natexlab{}.
\newblock \bibinfo{booktitle}{\emph{Automating inequality: How high-tech tools
  profile, police, and punish the poor}}.
\newblock \bibinfo{publisher}{St. Martin's Press}.
\newblock


\bibitem[\protect\citeauthoryear{for Fundamental Rights~(FRA)}{for Fundamental
  Rights~(FRA)}{line}]%
        {fraopinion}
\bibfield{author}{\bibinfo{person}{European Union~Agency for Fundamental
  Rights~(FRA)}.} \bibinfo{year}{2013 [Online]}\natexlab{}.
\newblock \bibinfo{title}{FRA Opinion on the situation of equality in the
  European Union 10 years on from initial implementation of the equality
  directives}.
\newblock
\newblock
\urldef\tempurl%
\url{https://fra.europa.eu/en/publication/2013/fra-opinion-situation-equality-european-union-10-years-initial-implementation}
\showURL{%
\tempurl}


\bibitem[\protect\citeauthoryear{for Fundamental Rights~(FRA) and
  of~Human~Rights}{for Fundamental Rights~(FRA) and of~Human~Rights}{2018}]%
        {nondiscriminationhb}
\bibfield{author}{\bibinfo{person}{European Union~Agency for Fundamental
  Rights~(FRA)} {and} \bibinfo{person}{European~Court of Human~Rights}.}
  \bibinfo{year}{2018}\natexlab{}.
\newblock \bibinfo{booktitle}{\emph{Handbook on European non-discrimination
  law}}.
\newblock \bibinfo{publisher}{Publications Office of the European Union},
  \bibinfo{address}{Luxembourg}.
\newblock


\bibitem[\protect\citeauthoryear{Gabriel}{Gabriel}{2020}]%
        {gabriel2020artificial}
\bibfield{author}{\bibinfo{person}{Iason Gabriel}.}
  \bibinfo{year}{2020}\natexlab{}.
\newblock \showarticletitle{Artificial intelligence, values, and alignment}.
\newblock \bibinfo{journal}{\emph{Minds and machines}} \bibinfo{volume}{30},
  \bibinfo{number}{3} (\bibinfo{year}{2020}), \bibinfo{pages}{411--437}.
\newblock


\bibitem[\protect\citeauthoryear{Gabriel}{Gabriel}{2021}]%
        {gabriel2021towards}
\bibfield{author}{\bibinfo{person}{Iason Gabriel}.}
  \bibinfo{year}{2021}\natexlab{}.
\newblock \showarticletitle{Towards a Theory of Justice for Artificial
  Intelligence}.
\newblock \bibinfo{journal}{\emph{arXiv preprint}} (\bibinfo{year}{2021}).
\newblock


\bibitem[\protect\citeauthoryear{Gerards and Xenidis}{Gerards and
  Xenidis}{2021}]%
        {gerards2021algorithmic}
\bibfield{author}{\bibinfo{person}{Janneke Gerards} {and}
  \bibinfo{person}{Raphaele Xenidis}.} \bibinfo{year}{2021}\natexlab{}.
\newblock \showarticletitle{Algorithmic Discrimination in Europe: Challenges
  and Opportunities for Gender Equality and Non-Discrimination Law}.
\newblock  (\bibinfo{year}{2021}).
\newblock


\bibitem[\protect\citeauthoryear{Gignac, Jetha, Ginis, and Ibrahim}{Gignac
  et~al\mbox{.}}{2021}]%
        {gignac2021does}
\bibfield{author}{\bibinfo{person}{Monique~AM Gignac}, \bibinfo{person}{Arif
  Jetha}, \bibinfo{person}{Kathleen A~Martin Ginis}, {and}
  \bibinfo{person}{Selahadin Ibrahim}.} \bibinfo{year}{2021}\natexlab{}.
\newblock \showarticletitle{Does it matter what your reasons are when deciding
  to disclose (or not disclose) a disability at work? The association of
  workers’ approach and avoidance goals with perceived positive and negative
  workplace outcomes}.
\newblock \bibinfo{journal}{\emph{Journal of Occupational Rehabilitation}}
  (\bibinfo{year}{2021}), \bibinfo{pages}{1--14}.
\newblock


\bibitem[\protect\citeauthoryear{Gugnani and Misra}{Gugnani and Misra}{2020}]%
        {gugnani2020implicit}
\bibfield{author}{\bibinfo{person}{Akshay Gugnani} {and}
  \bibinfo{person}{Hemant Misra}.} \bibinfo{year}{2020}\natexlab{}.
\newblock \showarticletitle{Implicit Skills Extraction Using Document Embedding
  and Its Use in Job Recommendation}. In \bibinfo{booktitle}{\emph{Proceedings
  of the AAAI Conference on Artificial Intelligence}},
  Vol.~\bibinfo{volume}{34}. \bibinfo{pages}{13286--13293}.
\newblock


\bibitem[\protect\citeauthoryear{Guo, Kamar, Vaughan, Wallach, and Morris}{Guo
  et~al\mbox{.}}{2020}]%
        {guo2020toward}
\bibfield{author}{\bibinfo{person}{Anhong Guo}, \bibinfo{person}{Ece Kamar},
  \bibinfo{person}{Jennifer~Wortman Vaughan}, \bibinfo{person}{Hanna Wallach},
  {and} \bibinfo{person}{Meredith~Ringel Morris}.}
  \bibinfo{year}{2020}\natexlab{}.
\newblock \showarticletitle{Toward fairness in AI for people with disabilities:
  a research roadmap}.
\newblock \bibinfo{journal}{\emph{ACM SIGACCESS Accessibility and Computing}}
  \bibinfo{number}{125} (\bibinfo{year}{2020}).
\newblock


\bibitem[\protect\citeauthoryear{Ha-Thuc, Venkataraman, Rodriguez, Sinha,
  Sundaram, and Guo}{Ha-Thuc et~al\mbox{.}}{2015}]%
        {ha2015personalized}
\bibfield{author}{\bibinfo{person}{Viet Ha-Thuc}, \bibinfo{person}{Ganesh
  Venkataraman}, \bibinfo{person}{Mario Rodriguez}, \bibinfo{person}{Shakti
  Sinha}, \bibinfo{person}{Senthil Sundaram}, {and} \bibinfo{person}{Lin Guo}.}
  \bibinfo{year}{2015}\natexlab{}.
\newblock \showarticletitle{Personalized expertise search at Linkedin}. In
  \bibinfo{booktitle}{\emph{IEEE International Conference on Big Data (Big
  Data)}}. IEEE, \bibinfo{pages}{1238--1247}.
\newblock


\bibitem[\protect\citeauthoryear{Hacker}{Hacker}{2018}]%
        {hacker2018teaching}
\bibfield{author}{\bibinfo{person}{Philipp Hacker}.}
  \bibinfo{year}{2018}\natexlab{}.
\newblock \showarticletitle{Teaching fairness to artificial intelligence:
  Existing and novel strategies against algorithmic discrimination under EU
  law}.
\newblock \bibinfo{journal}{\emph{Common Market Law Review}}
  \bibinfo{volume}{55}, \bibinfo{number}{4} (\bibinfo{year}{2018}).
\newblock


\bibitem[\protect\citeauthoryear{Hildebrandt}{Hildebrandt}{2020}]%
        {hildebrandt2020code}
\bibfield{author}{\bibinfo{person}{Mireille Hildebrandt}.}
  \bibinfo{year}{2020}\natexlab{}.
\newblock \showarticletitle{Code-driven Law: Freezing the Future and Scaling
  the Past}.
\newblock \bibinfo{journal}{\emph{Is Law Computable?: Critical Perspectives on
  Law and Artificial Intelligence}} (\bibinfo{year}{2020}),
  \bibinfo{pages}{67}.
\newblock


\bibitem[\protect\citeauthoryear{Hildebrandt}{Hildebrandt}{2021a}]%
        {hildebrandt2021discrimination}
\bibfield{author}{\bibinfo{person}{Mireille Hildebrandt}.}
  \bibinfo{year}{2021}\natexlab{a}.
\newblock \showarticletitle{Discrimination, Data-driven AI Systems and
  Practical Reason}.
\newblock \bibinfo{journal}{\emph{European Data Protection Law Review}}
  \bibinfo{volume}{7} (\bibinfo{year}{2021}), \bibinfo{pages}{358--366}.
\newblock


\bibitem[\protect\citeauthoryear{Hildebrandt}{Hildebrandt}{2021b}]%
        {hildebrandt2021issue}
\bibfield{author}{\bibinfo{person}{Mireille Hildebrandt}.}
  \bibinfo{year}{2021}\natexlab{b}.
\newblock \bibinfo{title}{The Issue of Proxies, And why EU law matters for
  recommender systems}.  (\bibinfo{year}{2021}).
\newblock
\urldef\tempurl%
\url{https://osf.io/preprints/socarxiv/45x67/}
\showURL{%
\tempurl}
\newblock
\shownote{(submitted for publication).}


\bibitem[\protect\citeauthoryear{Hutchinson, Prabhakaran, Denton, Webster,
  Zhong, and Denuyl}{Hutchinson et~al\mbox{.}}{2020}]%
        {hutchinson2020unintended}
\bibfield{author}{\bibinfo{person}{Ben Hutchinson}, \bibinfo{person}{Vinodkumar
  Prabhakaran}, \bibinfo{person}{Emily Denton}, \bibinfo{person}{Kellie
  Webster}, \bibinfo{person}{Yu Zhong}, {and} \bibinfo{person}{Stephen
  Denuyl}.} \bibinfo{year}{2020}\natexlab{}.
\newblock \showarticletitle{Unintended machine learning biases as social
  barriers for persons with disabilitiess}.
\newblock \bibinfo{journal}{\emph{ACM SIGACCESS Accessibility and Computing}}
  \bibinfo{number}{125} (\bibinfo{year}{2020}), \bibinfo{pages}{1--1}.
\newblock


\bibitem[\protect\citeauthoryear{Jobin, Ienca, and Vayena}{Jobin
  et~al\mbox{.}}{2019}]%
        {jobin2019global}
\bibfield{author}{\bibinfo{person}{Anna Jobin}, \bibinfo{person}{Marcello
  Ienca}, {and} \bibinfo{person}{Effy Vayena}.}
  \bibinfo{year}{2019}\natexlab{}.
\newblock \showarticletitle{The global landscape of AI ethics guidelines}.
\newblock \bibinfo{journal}{\emph{Nature Machine Intelligence}}
  \bibinfo{volume}{1}, \bibinfo{number}{9} (\bibinfo{year}{2019}),
  \bibinfo{pages}{389--399}.
\newblock


\bibitem[\protect\citeauthoryear{Kaminski and Malgieri}{Kaminski and
  Malgieri}{2020}]%
        {kaminski2020multi}
\bibfield{author}{\bibinfo{person}{Margot~E Kaminski} {and}
  \bibinfo{person}{Gianclaudio Malgieri}.} \bibinfo{year}{2020}\natexlab{}.
\newblock \showarticletitle{Multi-layered explanations from algorithmic impact
  assessments in the GDPR}. In \bibinfo{booktitle}{\emph{Proceedings of the
  2020 Conference on Fairness, Accountability, and Transparency}}.
  \bibinfo{pages}{68--79}.
\newblock


\bibitem[\protect\citeauthoryear{Kenthapadi, Le, and Venkataraman}{Kenthapadi
  et~al\mbox{.}}{2017}]%
        {kenthapadi2017personalized}
\bibfield{author}{\bibinfo{person}{Krishnaram Kenthapadi},
  \bibinfo{person}{Benjamin Le}, {and} \bibinfo{person}{Ganesh Venkataraman}.}
  \bibinfo{year}{2017}\natexlab{}.
\newblock \showarticletitle{Personalized job recommendation system at linkedin:
  Practical challenges and lessons learned}. In
  \bibinfo{booktitle}{\emph{Proceedings of the eleventh ACM conference on
  recommender systems}}. \bibinfo{pages}{346--347}.
\newblock


\bibitem[\protect\citeauthoryear{Kidzi{\'n}ski, Ong, Mohanty, Hicks, Carroll,
  Zhou, Zeng, Wang, Lian, Tian, et~al\mbox{.}}{Kidzi{\'n}ski
  et~al\mbox{.}}{2020}]%
        {kidzinski2020artificial}
\bibfield{author}{\bibinfo{person}{{\L}ukasz Kidzi{\'n}ski},
  \bibinfo{person}{Carmichael Ong}, \bibinfo{person}{Sharada~Prasanna Mohanty},
  \bibinfo{person}{Jennifer Hicks}, \bibinfo{person}{Sean Carroll},
  \bibinfo{person}{Bo Zhou}, \bibinfo{person}{Hongsheng Zeng},
  \bibinfo{person}{Fan Wang}, \bibinfo{person}{Rongzhong Lian},
  \bibinfo{person}{Hao Tian}, {et~al\mbox{.}}} \bibinfo{year}{2020}\natexlab{}.
\newblock \showarticletitle{Artificial intelligence for prosthetics: Challenge
  solutions}.
\newblock In \bibinfo{booktitle}{\emph{The NeurIPS'18 Competition}}.
  \bibinfo{publisher}{Springer}, \bibinfo{pages}{69--128}.
\newblock


\bibitem[\protect\citeauthoryear{Kim and Bodie}{Kim and Bodie}{2021}]%
        {kim2021artificial}
\bibfield{author}{\bibinfo{person}{Pauline~T Kim} {and}
  \bibinfo{person}{Matthew~T Bodie}.} \bibinfo{year}{2021}\natexlab{}.
\newblock \showarticletitle{Artificial Intelligence and the Challenges of
  Workplace Discrimination and Privacy}.
\newblock \bibinfo{journal}{\emph{Journal of Labor and Employment Law}}
  \bibinfo{volume}{35}, \bibinfo{number}{2} (\bibinfo{year}{2021}),
  \bibinfo{pages}{289--315}.
\newblock


\bibitem[\protect\citeauthoryear{Kroll, Huey, Barocas, Felten, Reidenberg,
  Robinson, and Yu}{Kroll et~al\mbox{.}}{2017}]%
        {krollaccountable}
\bibfield{author}{\bibinfo{person}{Joshua~A Kroll}, \bibinfo{person}{Joanna
  Huey}, \bibinfo{person}{Solon Barocas}, \bibinfo{person}{Edward~W Felten},
  \bibinfo{person}{Joel~R Reidenberg}, \bibinfo{person}{David~G Robinson},
  {and} \bibinfo{person}{Harlan Yu}.} \bibinfo{year}{2017}\natexlab{}.
\newblock \showarticletitle{Accountable Algorithms}.
\newblock \bibinfo{journal}{\emph{University of Pennsylvania Law Review}}
  \bibinfo{volume}{165} (\bibinfo{year}{2017}), \bibinfo{pages}{633--705}.
\newblock


\bibitem[\protect\citeauthoryear{Leo, Medioni, Trivedi, Kanade, and
  Farinella}{Leo et~al\mbox{.}}{2017}]%
        {leo2017computer}
\bibfield{author}{\bibinfo{person}{Marco Leo}, \bibinfo{person}{G Medioni},
  \bibinfo{person}{M Trivedi}, \bibinfo{person}{Takeo Kanade}, {and}
  \bibinfo{person}{Giovanni~Maria Farinella}.} \bibinfo{year}{2017}\natexlab{}.
\newblock \showarticletitle{Computer vision for assistive technologies}.
\newblock \bibinfo{journal}{\emph{Computer Vision and Image Understanding}}
  \bibinfo{volume}{154} (\bibinfo{year}{2017}), \bibinfo{pages}{1--15}.
\newblock


\bibitem[\protect\citeauthoryear{Lerman}{Lerman}{2013}]%
        {lerman2013big}
\bibfield{author}{\bibinfo{person}{Jonas Lerman}.}
  \bibinfo{year}{2013}\natexlab{}.
\newblock \showarticletitle{Big data and its exclusions}.
\newblock \bibinfo{journal}{\emph{Stan. L. Rev. Online}}  \bibinfo{volume}{66}
  (\bibinfo{year}{2013}), \bibinfo{pages}{55}.
\newblock


\bibitem[\protect\citeauthoryear{Lillywhite and Wolbring}{Lillywhite and
  Wolbring}{2020}]%
        {lillywhite2020coverage}
\bibfield{author}{\bibinfo{person}{Aspen Lillywhite} {and}
  \bibinfo{person}{Gregor Wolbring}.} \bibinfo{year}{2020}\natexlab{}.
\newblock \showarticletitle{Coverage of artificial intelligence and machine
  learning within academic literature, Canadian newspapers, and twitter tweets:
  The case of disabled people}.
\newblock \bibinfo{journal}{\emph{Societies}} \bibinfo{volume}{10},
  \bibinfo{number}{1} (\bibinfo{year}{2020}), \bibinfo{pages}{23}.
\newblock


\bibitem[\protect\citeauthoryear{Makkonen}{Makkonen}{2016}]%
        {europequalityhb}
\bibfield{author}{\bibinfo{person}{Timo Makkonen}.}
  \bibinfo{year}{2016}\natexlab{}.
\newblock \bibinfo{booktitle}{\emph{European Handbook on Equality Data}}.
\newblock \bibinfo{publisher}{Publications Office of the European Union},
  \bibinfo{address}{Luxembourg}.
\newblock
\urldef\tempurl%
\url{https://ec.europa.eu/info/sites/default/files/european_handbook_on_equality.pdf}
\showURL{%
\tempurl}


\bibitem[\protect\citeauthoryear{Malgieri and Comand{\'e}}{Malgieri and
  Comand{\'e}}{2017}]%
        {malgieri2017right}
\bibfield{author}{\bibinfo{person}{Gianclaudio Malgieri} {and}
  \bibinfo{person}{Giovanni Comand{\'e}}.} \bibinfo{year}{2017}\natexlab{}.
\newblock \showarticletitle{Why a right to legibility of automated
  decision-making exists in the general data protection regulation}.
\newblock \bibinfo{journal}{\emph{International Data Privacy Law}}
  (\bibinfo{year}{2017}).
\newblock


\bibitem[\protect\citeauthoryear{Mantelero}{Mantelero}{2017}]%
        {mantelero2017group}
\bibfield{author}{\bibinfo{person}{Alessandro Mantelero}.}
  \bibinfo{year}{2017}\natexlab{}.
\newblock \showarticletitle{From group privacy to collective privacy: towards a
  new dimension of privacy and data protection in the big data era}.
\newblock In \bibinfo{booktitle}{\emph{Group Privacy}}.
  \bibinfo{publisher}{Springer}, \bibinfo{pages}{139--158}.
\newblock


\bibitem[\protect\citeauthoryear{{Marie Lecerf}}{{Marie Lecerf}}{2020}]%
        {eupe651932}
\bibfield{author}{\bibinfo{person}{{Marie Lecerf}}.}
  \bibinfo{year}{2020}\natexlab{}.
\newblock \bibinfo{title}{Employment and disability in the European Union.
  Briefing PE 651.932.}
\newblock
  \bibinfo{howpublished}{\url{https://www.europarl.europa.eu/RegData/etudes/BRIE/2020/651932/EPRS_BRI(2020)651932_EN.pdf}}.
\newblock
\newblock
\shownote{Accessed: 2021-01-7.}


\bibitem[\protect\citeauthoryear{Mehrabi, Morstatter, Saxena, Lerman, and
  Galstyan}{Mehrabi et~al\mbox{.}}{2021}]%
        {mehrabi2021survey}
\bibfield{author}{\bibinfo{person}{Ninareh Mehrabi}, \bibinfo{person}{Fred
  Morstatter}, \bibinfo{person}{Nripsuta Saxena}, \bibinfo{person}{Kristina
  Lerman}, {and} \bibinfo{person}{Aram Galstyan}.}
  \bibinfo{year}{2021}\natexlab{}.
\newblock \showarticletitle{A survey on bias and fairness in machine learning}.
\newblock \bibinfo{journal}{\emph{ACM Computing Surveys (CSUR)}}
  \bibinfo{volume}{54}, \bibinfo{number}{6} (\bibinfo{year}{2021}),
  \bibinfo{pages}{1--35}.
\newblock


\bibitem[\protect\citeauthoryear{Mendoza and Bygrave}{Mendoza and
  Bygrave}{2017}]%
        {mendoza2017right}
\bibfield{author}{\bibinfo{person}{Isak Mendoza} {and} \bibinfo{person}{Lee~A
  Bygrave}.} \bibinfo{year}{2017}\natexlab{}.
\newblock \showarticletitle{The right not to be subject to automated decisions
  based on profiling}.
\newblock In \bibinfo{booktitle}{\emph{EU Internet Law}}.
  \bibinfo{publisher}{Springer}, \bibinfo{pages}{77--98}.
\newblock


\bibitem[\protect\citeauthoryear{Mostofa, Fullin, Zehtabian, Bacanl{\i},
  B{\"o}l{\"o}ni, and Turgut}{Mostofa et~al\mbox{.}}{2020}]%
        {mostofa2020iot}
\bibfield{author}{\bibinfo{person}{Nafisa Mostofa}, \bibinfo{person}{Kelly
  Fullin}, \bibinfo{person}{Sharare Zehtabian}, \bibinfo{person}{Safa
  Bacanl{\i}}, \bibinfo{person}{Ladislau B{\"o}l{\"o}ni}, {and}
  \bibinfo{person}{Damla Turgut}.} \bibinfo{year}{2020}\natexlab{}.
\newblock \showarticletitle{IoT-Enabled Smart Mobility Devices for Aging and
  Rehabilitation}. In \bibinfo{booktitle}{\emph{IEEE International Conference
  on Communications (ICC)}}. IEEE, \bibinfo{pages}{1--6}.
\newblock


\bibitem[\protect\citeauthoryear{Muir}{Muir}{2013}]%
        {muir2013transformative}
\bibfield{author}{\bibinfo{person}{Elise Muir}.}
  \bibinfo{year}{2013}\natexlab{}.
\newblock \showarticletitle{The transformative function of EU equality law}.
\newblock \bibinfo{journal}{\emph{European Review of Private Law}}
  \bibinfo{volume}{21}, \bibinfo{number}{5/6} (\bibinfo{year}{2013}),
  \bibinfo{pages}{1231--1253}.
\newblock


\bibitem[\protect\citeauthoryear{Muir}{Muir}{2018}]%
        {muir2018eu}
\bibfield{author}{\bibinfo{person}{Elise Muir}.}
  \bibinfo{year}{2018}\natexlab{}.
\newblock \bibinfo{booktitle}{\emph{EU Equality Law: The First Fundamental
  Rights Policy of the EU}}.
\newblock \bibinfo{publisher}{Oxford University Press}.
\newblock


\bibitem[\protect\citeauthoryear{Mulfari}{Mulfari}{2018}]%
        {mulfari2018tensorflow}
\bibfield{author}{\bibinfo{person}{Davide Mulfari}.}
  \bibinfo{year}{2018}\natexlab{}.
\newblock \showarticletitle{A TensorFlow-based assistive technology system for
  users with visual impairments}. In \bibinfo{booktitle}{\emph{Proceedings of
  the 15th International Web for All Conference}}. \bibinfo{pages}{1--2}.
\newblock


\bibitem[\protect\citeauthoryear{Network}{Network}{2018}]%
        {adannguideline}
\bibfield{author}{\bibinfo{person}{ADA~National Network}.}
  \bibinfo{year}{2018}\natexlab{}.
\newblock \bibinfo{title}{Guidelines for Writing About People With
  Disabilities}.
\newblock
  \bibinfo{howpublished}{\url{https://adata.org/factsheet/ADANN-writing}}.
\newblock
\newblock
\shownote{Accessed: 2021-01-7.}


\bibitem[\protect\citeauthoryear{Nugent, Jackson, Scott-Parker, Partridge,
  Raper, Bakalis, Shepherd, Mitra, Long, Maynard, et~al\mbox{.}}{Nugent
  et~al\mbox{.}}{2020}]%
        {nugent2020recruitment}
\bibfield{author}{\bibinfo{person}{Selin Nugent}, \bibinfo{person}{Paul
  Jackson}, \bibinfo{person}{Susan Scott-Parker}, \bibinfo{person}{James
  Partridge}, \bibinfo{person}{Rebecca Raper}, \bibinfo{person}{Chara Bakalis},
  \bibinfo{person}{Alex Shepherd}, \bibinfo{person}{Arijit Mitra},
  \bibinfo{person}{Jintao Long}, \bibinfo{person}{Kevin Maynard},
  {et~al\mbox{.}}} \bibinfo{year}{2020}\natexlab{}.
\newblock \showarticletitle{Recruitment AI has a Disability Problem: questions
  employers should be asking to ensure fairness in recruitment}.
\newblock  (\bibinfo{year}{2020}).
\newblock


\bibitem[\protect\citeauthoryear{of~the European~Union}{of~the
  European~Union}{2000}]%
        {EED2000}
\bibfield{author}{\bibinfo{person}{Council of~the European~Union}.}
  \bibinfo{year}{2000}\natexlab{}.
\newblock \showarticletitle{Council Directive 2000/78/EC of 27 November 2000
  establishing a general framework for equal treatment in employment and
  occupation}.
\newblock \bibinfo{journal}{\emph{Official Journal of the European Union}}
  \bibinfo{volume}{L 303} (\bibinfo{year}{2000}).
\newblock
\urldef\tempurl%
\url{https://eur-lex.europa.eu/legal-content/EN/TXT/?uri=celex\%3A32000L0078}
\showURL{%
\tempurl}


\bibitem[\protect\citeauthoryear{Olkin}{Olkin}{2002}]%
        {olkin2002could}
\bibfield{author}{\bibinfo{person}{Rhoda Olkin}.}
  \bibinfo{year}{2002}\natexlab{}.
\newblock \showarticletitle{Could you hold the door for me? Including
  disability in diversity.}
\newblock \bibinfo{journal}{\emph{Cultural Diversity and Ethnic Minority
  Psychology}} \bibinfo{volume}{8}, \bibinfo{number}{2} (\bibinfo{year}{2002}),
  \bibinfo{pages}{130}.
\newblock


\bibitem[\protect\citeauthoryear{on~the Rights of Persons~with
  Disabilities}{on~the Rights of Persons~with Disabilities}{2018}]%
        {UNcomment}
\bibfield{author}{\bibinfo{person}{Committee on~the Rights of Persons~with
  Disabilities}.} \bibinfo{year}{2018}\natexlab{}.
\newblock \showarticletitle{General comment no. 6 on equality and
  non-discrimination}. In \bibinfo{booktitle}{\emph{Proceedings of the 19th
  session, Geneva}}.
\newblock
\urldef\tempurl%
\url{https://digitallibrary.un.org/record/1626976}
\showURL{%
\tempurl}


\bibitem[\protect\citeauthoryear{O'neil}{O'neil}{2016}]%
        {o2016weapons}
\bibfield{author}{\bibinfo{person}{Cathy O'neil}.}
  \bibinfo{year}{2016}\natexlab{}.
\newblock \bibinfo{booktitle}{\emph{Weapons of math destruction: How big data
  increases inequality and threatens democracy}}.
\newblock \bibinfo{publisher}{Crown}, \bibinfo{address}{New York}.
\newblock


\bibitem[\protect\citeauthoryear{Parliament and Council}{Parliament and
  Council}{2016}]%
        {GDPR2016}
\bibfield{author}{\bibinfo{person}{European Parliament} {and}
  \bibinfo{person}{Council}.} \bibinfo{year}{2016}\natexlab{}.
\newblock \showarticletitle{Regulation (EU) 2016/679 of the European Parliament
  and of the Council of 27 April 2016 on the protection of natural persons with
  regard to the processing of personal data and on the free movement of such
  data, and repealing Directive 95/46/EC (General Data Protection Regulation)}.
\newblock \bibinfo{journal}{\emph{Official Journal of the European Union}}
  \bibinfo{volume}{L 119} (\bibinfo{year}{2016}), \bibinfo{pages}{1--88}.
\newblock
\urldef\tempurl%
\url{https://eur-lex.europa.eu/eli/reg/2016/679/oj}
\showURL{%
\tempurl}


\bibitem[\protect\citeauthoryear{Party}{Party}{2001}]%
        {WPopinion8}
\bibfield{author}{\bibinfo{person}{Data Protection~Working Party}.}
  \bibinfo{year}{2001}\natexlab{}.
\newblock \bibinfo{title}{Opinion 8/2001 on the processing of personal data in
  the employment context}.
\newblock \bibinfo{howpublished}{5062/01/EN/Final, WP 48}.
\newblock
\urldef\tempurl%
\url{https://ec.europa.eu/justice/article-29/documentation/opinion-recommendation/files/2001/wp48_en.pdf}
\showURL{%
\tempurl}


\bibitem[\protect\citeauthoryear{Party}{Party}{2017}]%
        {WPguidelinesDPIA}
\bibfield{author}{\bibinfo{person}{Data Protection~Working Party}.}
  \bibinfo{year}{2017}\natexlab{}.
\newblock \bibinfo{title}{Guidelines on Data Protection Impact Assessment
  (DPIA) and determining whether processing is “likely to result in a high
  risk” for the purposes of Regulation 2016/679}.
\newblock \bibinfo{howpublished}{WP 248 rev.01}.
\newblock
\urldef\tempurl%
\url{https://ec.europa.eu/newsroom/just/document.cfm?doc_id=47711}
\showURL{%
\tempurl}


\bibitem[\protect\citeauthoryear{Party}{Party}{2018}]%
        {WPguidelines}
\bibfield{author}{\bibinfo{person}{Data Protection~Working Party}.}
  \bibinfo{year}{2018}\natexlab{}.
\newblock \bibinfo{title}{Guidelines on Automated individual decision-making
  and Profiling for the purposes of Regulation 2016/679}.
\newblock \bibinfo{howpublished}{WP251rev.01}.
\newblock
\urldef\tempurl%
\url{https://ec.europa.eu/newsroom/article29/redirection/document/49826}
\showURL{%
\tempurl}


\bibitem[\protect\citeauthoryear{Pasquale}{Pasquale}{2015}]%
        {pasquale2015black}
\bibfield{author}{\bibinfo{person}{Frank Pasquale}.}
  \bibinfo{year}{2015}\natexlab{}.
\newblock \bibinfo{booktitle}{\emph{The black box society}}.
\newblock \bibinfo{publisher}{Harvard University Press}.
\newblock


\bibitem[\protect\citeauthoryear{Pedreshi, Ruggieri, and Turini}{Pedreshi
  et~al\mbox{.}}{2008}]%
        {pedreshi2008discrimination}
\bibfield{author}{\bibinfo{person}{Dino Pedreshi}, \bibinfo{person}{Salvatore
  Ruggieri}, {and} \bibinfo{person}{Franco Turini}.}
  \bibinfo{year}{2008}\natexlab{}.
\newblock \showarticletitle{Discrimination-aware data mining}. In
  \bibinfo{booktitle}{\emph{Proceedings of the 14th ACM SIGKDD international
  conference on Knowledge discovery and data mining}}.
  \bibinfo{pages}{560--568}.
\newblock


\bibitem[\protect\citeauthoryear{Persson and Hansson}{Persson and
  Hansson}{2003}]%
        {persson2003privacy}
\bibfield{author}{\bibinfo{person}{Anders~J Persson} {and}
  \bibinfo{person}{Sven~Ove Hansson}.} \bibinfo{year}{2003}\natexlab{}.
\newblock \showarticletitle{Privacy at work--ethical criteria}.
\newblock \bibinfo{journal}{\emph{Journal of Business Ethics}}
  \bibinfo{volume}{42}, \bibinfo{number}{1} (\bibinfo{year}{2003}),
  \bibinfo{pages}{59--70}.
\newblock


\bibitem[\protect\citeauthoryear{QC and Masters}{QC and Masters}{line}]%
        {equinet2020}
\bibfield{author}{\bibinfo{person}{Robin~Allen QC} {and} \bibinfo{person}{Dee
  Masters}.} \bibinfo{year}{2020 [Online]}\natexlab{}.
\newblock \bibinfo{title}{Regulating for an Equal AI: A new Role for Equality
  Bodies. Meeting the new Challenges to Equality and Non-Discrimination from
  Increased Digitisation and the Use of Artificial Intelligence}.
\newblock \bibinfo{howpublished}{An Equinet Publication}.
\newblock
\urldef\tempurl%
\url{https://equineteurope.org/wp-content/uploads/2020/06/ai_report_digital.pdf}
\showURL{%
\tempurl}


\bibitem[\protect\citeauthoryear{Quelle}{Quelle}{2018}]%
        {quelle2018enhancing}
\bibfield{author}{\bibinfo{person}{Claudia Quelle}.}
  \bibinfo{year}{2018}\natexlab{}.
\newblock \showarticletitle{Enhancing compliance under the general data
  protection regulation: The risky upshot of the accountability-and risk-based
  approach}.
\newblock \bibinfo{journal}{\emph{European Journal of Risk Regulation}}
  \bibinfo{volume}{9}, \bibinfo{number}{3} (\bibinfo{year}{2018}),
  \bibinfo{pages}{502--526}.
\newblock


\bibitem[\protect\citeauthoryear{Ramanath, Inan, Polatkan, Hu, Guo, Ozcaglar,
  Wu, Kenthapadi, and Geyik}{Ramanath et~al\mbox{.}}{2018}]%
        {ramanath2018towards}
\bibfield{author}{\bibinfo{person}{Rohan Ramanath}, \bibinfo{person}{Hakan
  Inan}, \bibinfo{person}{Gungor Polatkan}, \bibinfo{person}{Bo Hu},
  \bibinfo{person}{Qi Guo}, \bibinfo{person}{Cagri Ozcaglar},
  \bibinfo{person}{Xianren Wu}, \bibinfo{person}{Krishnaram Kenthapadi}, {and}
  \bibinfo{person}{Sahin~Cem Geyik}.} \bibinfo{year}{2018}\natexlab{}.
\newblock \showarticletitle{Towards deep and representation learning for talent
  search at LinkedIn}. In \bibinfo{booktitle}{\emph{Proceedings of the 27th ACM
  International Conference on Information and Knowledge Management}}.
  \bibinfo{pages}{2253--2261}.
\newblock


\bibitem[\protect\citeauthoryear{Rawls}{Rawls}{1971}]%
        {rawls1971theory}
\bibfield{author}{\bibinfo{person}{John Rawls}.}
  \bibinfo{year}{1971}\natexlab{}.
\newblock \bibinfo{booktitle}{\emph{A theory of justice}}.
\newblock \bibinfo{publisher}{Harvard University Press}.
\newblock


\bibitem[\protect\citeauthoryear{S{\'a}nchez-Monedero, Dencik, and
  Edwards}{S{\'a}nchez-Monedero et~al\mbox{.}}{2020}]%
        {sanchez2020does}
\bibfield{author}{\bibinfo{person}{Javier S{\'a}nchez-Monedero},
  \bibinfo{person}{Lina Dencik}, {and} \bibinfo{person}{Lilian Edwards}.}
  \bibinfo{year}{2020}\natexlab{}.
\newblock \showarticletitle{What does it mean to 'solve' the problem of
  discrimination in hiring? Social, technical and legal perspectives from the
  UK on automated hiring systems}. In \bibinfo{booktitle}{\emph{Proceedings of
  the 2020 conference on fairness, accountability, and transparency}}.
  \bibinfo{pages}{458--468}.
\newblock


\bibitem[\protect\citeauthoryear{Scherer and Shetty}{Scherer and
  Shetty}{2021}]%
        {scherer_shetty_2021}
\bibfield{author}{\bibinfo{person}{Matt Scherer} {and} \bibinfo{person}{Ridhi
  Shetty}.} \bibinfo{year}{2021}\natexlab{}.
\newblock \bibinfo{title}{NYC draft bill on AI in hiring needs higher and
  clearer hurdles}.
\newblock
\newblock
\urldef\tempurl%
\url{https://cdt.org/insights/nyc-draft-bill-on-ai-in-hiring-needs-higher-and-clearer-hurdles/}
\showURL{%
\tempurl}


\bibitem[\protect\citeauthoryear{Selbst and Powles}{Selbst and Powles}{2018}]%
        {selbst2018meaningful}
\bibfield{author}{\bibinfo{person}{Andrew Selbst} {and} \bibinfo{person}{Julia
  Powles}.} \bibinfo{year}{2018}\natexlab{}.
\newblock \showarticletitle{Meaningful Information and the Right to
  Explanation}. In \bibinfo{booktitle}{\emph{Conference on Fairness,
  Accountability and Transparency}}. PMLR, \bibinfo{pages}{48--48}.
\newblock


\bibitem[\protect\citeauthoryear{Selbst, Boyd, Friedler, Venkatasubramanian,
  and Vertesi}{Selbst et~al\mbox{.}}{2019}]%
        {selbst2019fairness}
\bibfield{author}{\bibinfo{person}{Andrew~D Selbst}, \bibinfo{person}{Danah
  Boyd}, \bibinfo{person}{Sorelle~A Friedler}, \bibinfo{person}{Suresh
  Venkatasubramanian}, {and} \bibinfo{person}{Janet Vertesi}.}
  \bibinfo{year}{2019}\natexlab{}.
\newblock \showarticletitle{Fairness and abstraction in sociotechnical
  systems}. In \bibinfo{booktitle}{\emph{Proceedings of the conference on
  fairness, accountability, and transparency}}. \bibinfo{pages}{59--68}.
\newblock


\bibitem[\protect\citeauthoryear{Smuha}{Smuha}{2021}]%
        {smuha2021beyond}
\bibfield{author}{\bibinfo{person}{Nathalie~A Smuha}.}
  \bibinfo{year}{2021}\natexlab{}.
\newblock \showarticletitle{Beyond a human rights-based approach to AI
  governance: Promise, pitfalls, plea}.
\newblock \bibinfo{journal}{\emph{Philosophy \& Technology}}
  \bibinfo{volume}{34}, \bibinfo{number}{1} (\bibinfo{year}{2021}),
  \bibinfo{pages}{91--104}.
\newblock


\bibitem[\protect\citeauthoryear{Smuha, Ahmed-Rengers, Harkens, Li, MacLaren,
  Piselli, and Yeung}{Smuha et~al\mbox{.}}{2021}]%
        {smuha2021eu}
\bibfield{author}{\bibinfo{person}{Nathalie~A Smuha}, \bibinfo{person}{Emma
  Ahmed-Rengers}, \bibinfo{person}{Adam Harkens}, \bibinfo{person}{Wenlong Li},
  \bibinfo{person}{James MacLaren}, \bibinfo{person}{Riccardo Piselli}, {and}
  \bibinfo{person}{Karen Yeung}.} \bibinfo{year}{2021}\natexlab{}.
\newblock \showarticletitle{How the EU Can Achieve Legally Trustworthy AI: A
  Response to the European Commission’s Proposal for an Artificial
  Intelligence Act}.
\newblock \bibinfo{journal}{\emph{SSRN}} (\bibinfo{year}{2021}).
\newblock


\bibitem[\protect\citeauthoryear{Snow}{Snow}{2004}]%
        {moralimperative}
\bibfield{author}{\bibinfo{person}{Kathie Snow}.}
  \bibinfo{year}{2004}\natexlab{}.
\newblock \bibinfo{title}{The Moral Imperative of Inclusion}.
\newblock
  \bibinfo{howpublished}{\url{https://nebula.wsimg.com/f091343f852ba6a73471bcf8d76e6f88?AccessKeyId=9D6F6082FE5EE52C3DC6}}.
\newblock
\newblock
\shownote{Accessed: 2021-01-7.}


\bibitem[\protect\citeauthoryear{Stangl, Morris, and Gurari}{Stangl
  et~al\mbox{.}}{2020}]%
        {stangl2020person}
\bibfield{author}{\bibinfo{person}{Abigale Stangl},
  \bibinfo{person}{Meredith~Ringel Morris}, {and} \bibinfo{person}{Danna
  Gurari}.} \bibinfo{year}{2020}\natexlab{}.
\newblock \showarticletitle{"Person, Shoes, Tree. Is the Person Naked?" What
  People with Vision Impairments Want in Image Descriptions}. In
  \bibinfo{booktitle}{\emph{Proceedings of the 2020 CHI Conference on Human
  Factors in Computing Systems}}. \bibinfo{pages}{1--13}.
\newblock


\bibitem[\protect\citeauthoryear{Trewin}{Trewin}{2018}]%
        {trewin2018ai}
\bibfield{author}{\bibinfo{person}{Shari Trewin}.}
  \bibinfo{year}{2018}\natexlab{}.
\newblock \showarticletitle{AI fairness for people with disabilities: Point of
  view}.
\newblock \bibinfo{journal}{\emph{arXiv preprint}} (\bibinfo{year}{2018}).
\newblock


\bibitem[\protect\citeauthoryear{Trewin, Basson, Muller, Branham, Treviranus,
  Gruen, Hebert, Lyckowski, and Manser}{Trewin et~al\mbox{.}}{2019}]%
        {trewin2019considerations}
\bibfield{author}{\bibinfo{person}{Shari Trewin}, \bibinfo{person}{Sara
  Basson}, \bibinfo{person}{Michael Muller}, \bibinfo{person}{Stacy Branham},
  \bibinfo{person}{Jutta Treviranus}, \bibinfo{person}{Daniel Gruen},
  \bibinfo{person}{Daniel Hebert}, \bibinfo{person}{Natalia Lyckowski}, {and}
  \bibinfo{person}{Erich Manser}.} \bibinfo{year}{2019}\natexlab{}.
\newblock \showarticletitle{Considerations for AI fairness for people with
  disabilities}.
\newblock \bibinfo{journal}{\emph{AI Matters}} \bibinfo{volume}{5},
  \bibinfo{number}{3} (\bibinfo{year}{2019}), \bibinfo{pages}{40--63}.
\newblock


\bibitem[\protect\citeauthoryear{van Bekkum and Borgesius}{van Bekkum and
  Borgesius}{2021}]%
        {specialcategory}
\bibfield{author}{\bibinfo{person}{Marvin van Bekkum} {and}
  \bibinfo{person}{Frederik~Zuiderveen Borgesius}.}
  \bibinfo{year}{2021}\natexlab{}.
\newblock \bibinfo{title}{Using special category data to prevent
  discrimination: does the GDPR need a new exception? Digital Legal Talks by
  Tilburg University}.  (\bibinfo{year}{2021}).
\newblock
\urldef\tempurl%
\url{https://www.sectorplandls.nl/wordpress/wp-content/uploads/2021/11/DLT-2021_paper_35.pdf}
\showURL{%
\tempurl}
\newblock
\shownote{Unpublished paper.}


\bibitem[\protect\citeauthoryear{Van~Caeneghem}{Van~Caeneghem}{2019}]%
        {van2019ethnic}
\bibfield{author}{\bibinfo{person}{Jozefien Van~Caeneghem}.}
  \bibinfo{year}{2019}\natexlab{}.
\newblock \showarticletitle{Ethnic Data Collection: Key Elements, Rules and
  Principles}.
\newblock In \bibinfo{booktitle}{\emph{Legal Aspects of Ethnic Data Collection
  and Positive Action}}. \bibinfo{publisher}{Springer},
  \bibinfo{pages}{155--257}.
\newblock


\bibitem[\protect\citeauthoryear{Veale and Borgesius}{Veale and
  Borgesius}{2021}]%
        {veale2021demystifying}
\bibfield{author}{\bibinfo{person}{Michael Veale} {and}
  \bibinfo{person}{Frederik~Zuiderveen Borgesius}.}
  \bibinfo{year}{2021}\natexlab{}.
\newblock \showarticletitle{Demystifying the Draft EU Artificial Intelligence
  Act—Analysing the good, the bad, and the unclear elements of the proposed
  approach}.
\newblock \bibinfo{journal}{\emph{Computer Law Review International}}
  \bibinfo{volume}{22}, \bibinfo{number}{4} (\bibinfo{year}{2021}),
  \bibinfo{pages}{97--112}.
\newblock


\bibitem[\protect\citeauthoryear{Wachter}{Wachter}{2021}]%
        {wachter2021richer}
\bibfield{author}{\bibinfo{person}{Sandra Wachter}.}
  \bibinfo{year}{2021}\natexlab{}.
\newblock \showarticletitle{How Fair AI Can Make Us Richer}.
\newblock \bibinfo{journal}{\emph{European Data Protection Law Review}}
  \bibinfo{volume}{7}, \bibinfo{number}{3} (\bibinfo{year}{2021}),
  \bibinfo{pages}{367--372}.
\newblock
\urldef\tempurl%
\url{https://doi.org/10.21552/edpl/2021/3/5}
\showURL{%
\tempurl}


\bibitem[\protect\citeauthoryear{Wachter, Mittelstadt, and Floridi}{Wachter
  et~al\mbox{.}}{2017}]%
        {wachter2017right}
\bibfield{author}{\bibinfo{person}{Sandra Wachter}, \bibinfo{person}{Brent
  Mittelstadt}, {and} \bibinfo{person}{Luciano Floridi}.}
  \bibinfo{year}{2017}\natexlab{}.
\newblock \showarticletitle{Why a right to explanation of automated
  decision-making does not exist in the general data protection regulation}.
\newblock \bibinfo{journal}{\emph{International Data Privacy Law}}
  \bibinfo{volume}{7}, \bibinfo{number}{2} (\bibinfo{year}{2017}),
  \bibinfo{pages}{76--99}.
\newblock


\bibitem[\protect\citeauthoryear{Wachter, Mittelstadt, and Russell}{Wachter
  et~al\mbox{.}}{2020}]%
        {wachter2020bias}
\bibfield{author}{\bibinfo{person}{Sandra Wachter}, \bibinfo{person}{Brent
  Mittelstadt}, {and} \bibinfo{person}{Chris Russell}.}
  \bibinfo{year}{2020}\natexlab{}.
\newblock \showarticletitle{Bias preservation in machine learning: the legality
  of fairness metrics under EU non-discrimination law}.
\newblock \bibinfo{journal}{\emph{W. Va. L. Rev.}}  \bibinfo{volume}{123}
  (\bibinfo{year}{2020}), \bibinfo{pages}{735}.
\newblock


\bibitem[\protect\citeauthoryear{Wachter, Mittelstadt, and Russell}{Wachter
  et~al\mbox{.}}{2021}]%
        {wachter2021fairness}
\bibfield{author}{\bibinfo{person}{Sandra Wachter}, \bibinfo{person}{Brent
  Mittelstadt}, {and} \bibinfo{person}{Chris Russell}.}
  \bibinfo{year}{2021}\natexlab{}.
\newblock \showarticletitle{Why fairness cannot be automated: Bridging the gap
  between EU non-discrimination law and AI}.
\newblock \bibinfo{journal}{\emph{Computer Law \& Security Review}}
  \bibinfo{volume}{41} (\bibinfo{year}{2021}).
\newblock


\bibitem[\protect\citeauthoryear{Whittaker, Alper, Bennett, Hendren, Kaziunas,
  Mills, Morris, Rankin, Rogers, Salas, et~al\mbox{.}}{Whittaker
  et~al\mbox{.}}{2019}]%
        {whittaker2019disability}
\bibfield{author}{\bibinfo{person}{Meredith Whittaker}, \bibinfo{person}{Meryl
  Alper}, \bibinfo{person}{Cynthia~L Bennett}, \bibinfo{person}{Sara Hendren},
  \bibinfo{person}{Liz Kaziunas}, \bibinfo{person}{Mara Mills},
  \bibinfo{person}{Meredith~Ringel Morris}, \bibinfo{person}{Joy Rankin},
  \bibinfo{person}{Emily Rogers}, \bibinfo{person}{Marcel Salas},
  {et~al\mbox{.}}} \bibinfo{year}{2019}\natexlab{}.
\newblock \showarticletitle{Disability, bias, and AI}.
\newblock \bibinfo{journal}{\emph{AI Now Institute}} (\bibinfo{year}{2019}).
\newblock


\bibitem[\protect\citeauthoryear{Wolff}{Wolff}{2021}]%
        {wolff2021technology}
\bibfield{author}{\bibinfo{person}{Josephine Wolff}.}
  \bibinfo{year}{2021}\natexlab{}.
\newblock \showarticletitle{How Is Technology Changing the World, and How
  Should the World Change Technology?}
\newblock \bibinfo{journal}{\emph{Global Perspectives}} \bibinfo{volume}{2},
  \bibinfo{number}{1} (\bibinfo{year}{2021}).
\newblock


\bibitem[\protect\citeauthoryear{{\v{Z}}liobait{\.e} and
  Custers}{{\v{Z}}liobait{\.e} and Custers}{2016}]%
        {vzliobaite2016using}
\bibfield{author}{\bibinfo{person}{Indr{\.e} {\v{Z}}liobait{\.e}} {and}
  \bibinfo{person}{Bart Custers}.} \bibinfo{year}{2016}\natexlab{}.
\newblock \showarticletitle{Using sensitive personal data may be necessary for
  avoiding discrimination in data-driven decision models}.
\newblock \bibinfo{journal}{\emph{Artificial Intelligence and Law}}
  \bibinfo{volume}{24}, \bibinfo{number}{2} (\bibinfo{year}{2016}),
  \bibinfo{pages}{183--201}.
\newblock


\end{thebibliography}

\end{document}